# $J_{\mathrm{eff}}$ = 1/2 Diamond Magnet $CaCo_2TeO_6$: Reimagining Frontiers of Spin Liquids and Quantum Functions

Xudong Huai,[1] Luke Pritchard Cairns,[2] Bridget Delles,[1] Michał J. Winiarski,[3] Maurice Sorolla II,[4] Xinshu Zhang,[5] Youzhe Chen,[2] Stuart Calder,[6] Eun Sang Choi,[7] Tatenda Kanyowa,[8] Anshul Kogar, [5] Huibo Cao,[6,8] Danielle Yahne,[6] Robert Birgeneau,[2] James Analytis,[2] Thao T. Tran*[1]

**Affiliations:**

[1]Department of Chemistry, Clemson University, Clemson, South Carolina 29634, United States

[2]Department of Physics, University of California, Berkeley, California, 94720, United States

[3]Faculty of Applied Physics and Mathematics and Advanced Materials Center, Gdansk University of Technology, Narutowicza 11/12, 80-233 Gdansk, Poland

[4]Institute of Chemistry, University of the Philippines Diliman, Quezon City 1101, Philippines

[5]Department of Physics and Astronomy, University of California, Los Angeles, California, 90095, United States

[6]Neutron Scattering Division, Oak Ridge National Laboratory, Oak Ridge, Tennessee 37830, United States

[7]National High Magnetic Field Laboratory, Tallahassee, Florida 32310, United States

[8]Materials Science and Engineering Department, University of Tennessee, Knoxville, Tennessee 37996, United States

**Abstract:** Diamond-lattice magnets, characterized by a framework of corner-sharing tetrahedra of magnetic cations, offer compelling avenues for realizing novel states of matter, with potential utilities in quantum technologies. However, focusing exclusively on spinels with $T_d$-site magnetic ions limits the tunability of competing Heisenberg interactions at comparable energy scales, hindering the ability to harness many-body electronic states and making the realization of $J_{\mathrm{eff}}$ = 1/2 and quantum spin liquids less accessible. We realize $CaCo_2TeO_6$—a new $J_{\mathrm{eff}}$ = 1/2 diamond magnet featuring $O_h$-$Co^{2+}$. This material displays enhanced competing Heisenberg exchange interactions, strong quantum fluctuations, and a field-induced quantum spin disordered state above 20 T. The results highlight the tantalizing physical and functional aspects of $CaCo_2TeO_6$, enabling a new pathway toward spin liquids and emergent quantum functionalities.

## Introduction

Spin liquids—quantum phases of matter—offer a unique array of features such as topology, fractionalization, and quantum entanglement. To unlock their full potential for implications in information science and technology, there is a need for innovative approaches to effectively engineer the energy levels of many-body electronic states in quantum systems.[1-6]

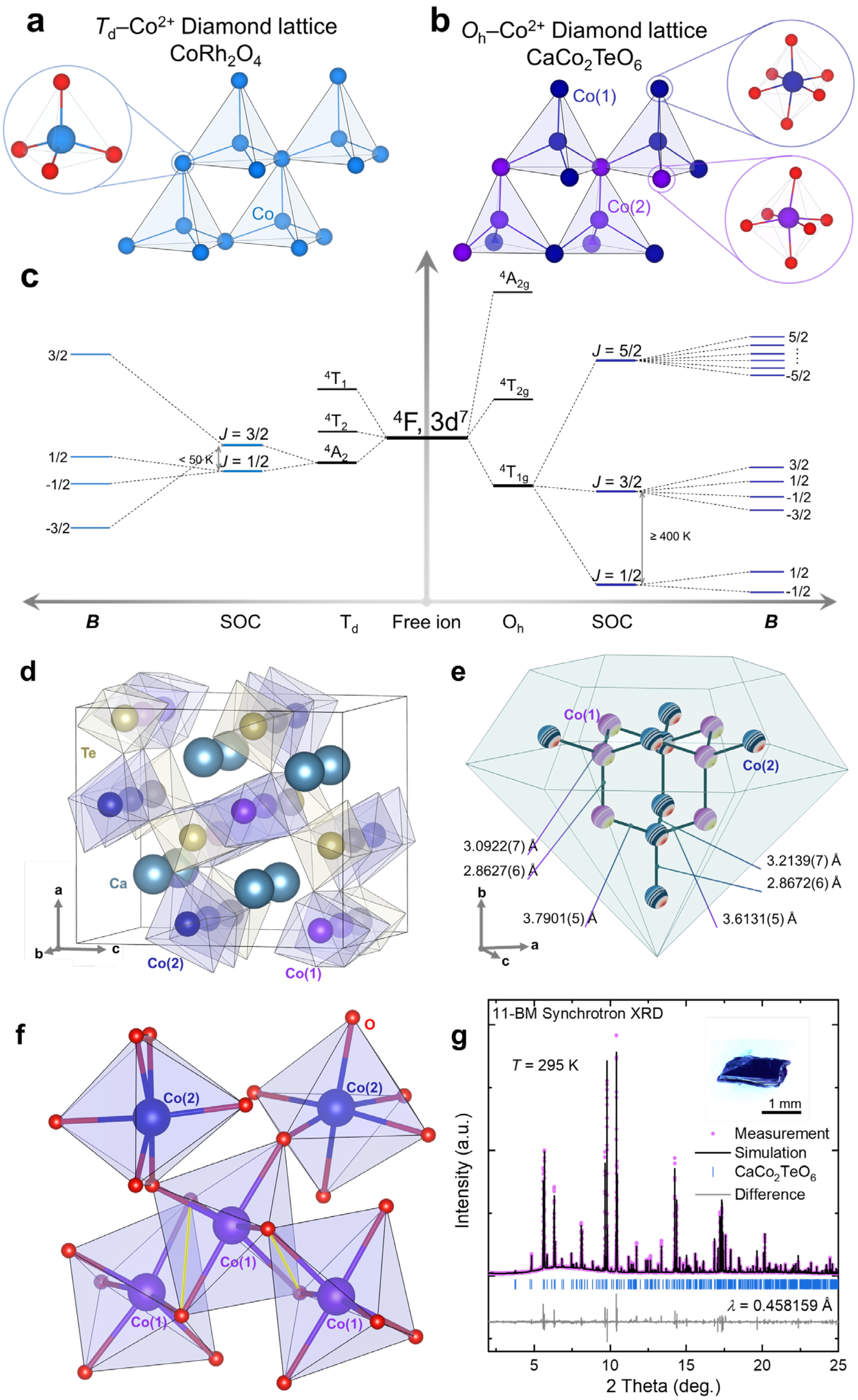

**Fig. 1** Diamond lattice of (a) $CoRh_2O_4$ with $T_d$-$Co^{2+}$ and (b) $CaCo_2TeO_6$ with $O_h$-$Co^{2+}$. (c) Energy level diagram of $T_d$ and $O_h$ crystal field under magnetic fields. (d) Crystal structure of $CaCo_2Te_6$. (e) Magnetic sublattice of $Co^{2+}$, showing different Co-Co distances. (f) View showing magnetic exchange pathways through edge-sharing and corner-sharing $[CoO_6]$ octahedra. (g) Rietveld fit (black) of high-resolution synchrotron XRD data (purple) of $CaCo_2TeO_6$, with the difference curve (gray) between experimental and calculated patterns.

Diamond-lattice Heisenberg antiferromagnets, with their distinctive 3-D framework of corner-sharing tetrahedra of magnetic cations, provide a fascinating space for realizing novel states of matter. Elegant studies on diamond-lattice magnets have mainly focused on spinels, where the magnetic site is placed in a tetrahedral ligand field ($T_d$) (Fig. 1a).[7-12] Among spinels, $T_d$-$Co^{2+}$ systems represent an exciting class of diamond magnets attributable to the potential to realize a $J_{eff}$ = 1/2 ground state and a quantum disordered state. [13-18] However, to the best of our knowledge, no such realizations have been experimentally demonstrated. A long-standing challenge in diamond magnets research has been the limited ability to modify the energy landscape of many-body electronic states and the competing Heisenberg exchange interactions $H = \sum_{i,j} J_{ij} S_i \cdot S_j$, both of which are crucial for achieving $J_{eff}$ = 1/2. This limitation has hindered the realization of quantum spin liquids and quantum functions at elevated temperatures (Fig. 1c). [19-25]

In this work, we present the case of a new diamond magnet—$CaCo_2TeO_6$—where, unlike $T_d$-$Co^{2+}$ spinels, the octahedral ligand fields of two distinct $O_h$-$Co^{2+}$ sites give rise to unique quantum properties with a $J_{eff}$ = 1/2 ground state (Fig. 1). Using a comprehensive approach that integrates experiments, characterization, and density functional theory (DFT) calculations, we demonstrate that the $O_h$-$Co^{2+}$ diamond magnet provides a powerful framework for fine-tuning both nearest-neighbor and next-nearest-neighbor Heisenberg interactions to comparable energy scales. This tunability facilitates the realization of a quantum spin disordered state above 20 T.

**Results and discussion**

$CaCo_2TeO_6$ displays a 3-D diamond lattice of $Co^{2+}$ with two crystallographically distinct Co(1) and Co(2) sites in a nearly perfect $O_h$ ligand field (Fig. 1). The $O_h$ local symmetry and the $^4T_{1g}$ ground state of the $Co^{2+}$ ($3d^7$) ion are consistent with the electronic transitions observed in spectroscopy measurements (Fig. S2-S3). The nearest Co–Co distances range from 2.8627(6) to 3.7901(5) Å (Fig. 1e), shorter than or comparable to that in a conventional diamond lattice (3.682 (2) Å in $CoRh_2O_4$).[12] The Co–Co separation variation originates from different fashions through which the $CoO_6$ octahedra are connected. Edge-sharing results in shorter Co–Co, while corner-sharing leads to Co–Co distances comparable to those in $CoRh_2O_4$, enabling the modulation of nearest- and next-nearest-neighbor Heisenberg exchange interactions (Fig. 1f).

$CaCo_2TeO_6$ undergoes antiferromagnetic (AFM) transitions at $T_{N1}$ = 16.5 K and $T_{N2}$ = 14.5 K at $\mu_0H$ = 0.01 T, clearly revealed by two anomalies in d$M$/d$T$ (Fig. 2a-c). Above $T_N$, the inverse magnetic susceptibility is well-described by the Curie-Weiss law for the paramagnetic regime. Curie-Weiss temperatures $\theta_{CW}$ of -76.3(1) K ($\mu_0H \parallel a$), -32.9(1) K ($\mu_0H \parallel b$), and -36.2(1) K ($\mu_0H \parallel c$) indicate a net AFM interaction and appreciable magnetic anisotropy. The effective magnetic moment $\mu_{eff}$ per $Co^{2+}$ ion is $\mu_{eff}$ = 4.96(1), 5.04(1), and 4.97(1) $\mu_B$ with $\mu_0H \parallel a$, $b$, and $c$ axis, respectively, closer to the expected value 4.74 $\mu_B$ ($S$ = 3/2 and $L$ = 1) than 5.67 $\mu_B$ ($S$ = 3/2 and $L$ = 2). The result agrees with a $J_{eff}$ = 1/2 ground state owing to the Kramers doublet.[22,26] The effective $g$-factor was extracted from the Curie constant to be $g_{eff}$ = 2.77 (Fig. S4a), greater than the electron-spin-only value $g_e$ = 2.0023, indicating a significant orbital contribution. At 400–700 K, the effective magnetic moment extracted from the Curie-Weiss analysis is $\mu_{eff}$ = 5.0(3) $\mu_B$ per $Co^{2+}$ (Fig. S4c-d), greater than the observed magnetic moment ~4.2 $\mu_B$ for the $J_{eff}$ = 1/2 ground state in $BaCo_3(VO_4)_2(OH)_2$ with $O_h$-$Co^{2+}$ kagome.[27] The increased moment in $CaCo_2TeO_6$ likely originates from its enhanced orbital contribution compared to that in $BaCo_3(VO_4)_2(OH)_2$, as substantiated in the aforementioned magnetization analysis and the $g_{eff}$ extraction. It is worth noting that $O_h$-$Co^{2+}$ systems typically enter a $J_{eff}$ = 3/2 level at ~150–300 K above $J_{eff}$ = 1/2,[28] nevertheless, the $J_{eff}$ = 1/2 ground state in $CaCo_2TeO_6$ is stabilized over a wide temperature range up to at least 400 K and well-separated from $J_{eff}$ = 3/2. This unique property of $CaCo_2TeO_6$ also sets this system apart from $T_d$-$Co^{2+}$ diamond magnets, wherein the separation between $J_{eff}$ = 1/2 and $J_{eff}$ = 3/2 is approximately 50 K.[29]

Fig. 2b-c depicts how the magnetic susceptibility of the material evolves under different fields near $T_N$. The magnetic susceptibility at low fields shows a transition at $T_{N1}$ = 16.5 K, followed by an AFM transition at $T_{N2}$ = 14.5 K. Upturns and downturns in $\chi(T)$ suggest competing FM and AFM interactions.[30] As magnetic fields increase, the two magnetic transitions are suppressed to lower temperatures and then evolve into only one transition observed above 4 T. Under higher fields, the transition is further suppressed to lower temperatures and eventually vanishes at $\mu_0H \geq 20$ T, revealing that the system enters a quantum disordered state (Fig S5).[31,32] This behavior is also clearly shown in d*M*/d*T* (Fig S5a), in which ridges can imply field-induced transitions. Below 12 K, upturns in $\chi(T)$ can be attributed to competing nearest-neighbor and next-nearest-neighbor Co–Co interactions, similar to that in $T_d$-$Co^{2+}$ diamond magnets.[12,33] However, the competing interactions of the $O_h$-$Co^{2+}$ spins are more complex than $T_d$-$Co^{2+}$, since there are six nearest-neighbor Co–Co interactions through Co(1)-Co(1), Co(2)-Co(2), and Co(1)-Co(2) (Table S2). Orientation-dependent isothermal magnetization *M*(*H*) curves display significant magnetic anisotropy (Fig. S6), which decreases with increasing temperature. *M*(*H*) curves with $\mu_0H /\!/ a$ show linear correlation at $\mu_0H \leq 10$ T, while those with $\mu_0H /\!/ b$ and *c* feature hysteresis loops below $T_{N1,2}$. The hysteresis loops feature zero coercivity and dumbbell shapes (Fig. 2d-f). While the net magnetic interactions in $CaCo_2TeO_6$ are dominated by AFM exchange, the double hysteresis loops indicate FM correlation. At 16 K, the hysteresis corresponding to FM exchange vanishes, but metamagnetic transitions are still observed. Similar double hysteresis loops have been observed in $BaCo_2(AsO_4)_2$ honeycomb.[34] As the applied field increases up to 35 T, the magnetic moment continues to rise without reaching saturation (Fig. S6). The moment exceeds 1.385 $\mu_B$ ($g_{eff}$ = 2.77, $J$ = 1/2), yet remains well below 3 $\mu_B$ ($g_e$ = 2, $J$ = 3/2), suggesting *J*-state mixing between the $J_{eff}$ = 1/2 and $J_{eff}$ = 3/2 states. Zeeman splitting in applied fields breaks the degeneracy within the $J_{eff}$ = 3/2 manifold, effectively reducing its energy gap with the $J_{eff}$ = ½ state and thus facilitating the degree of mixing between them.

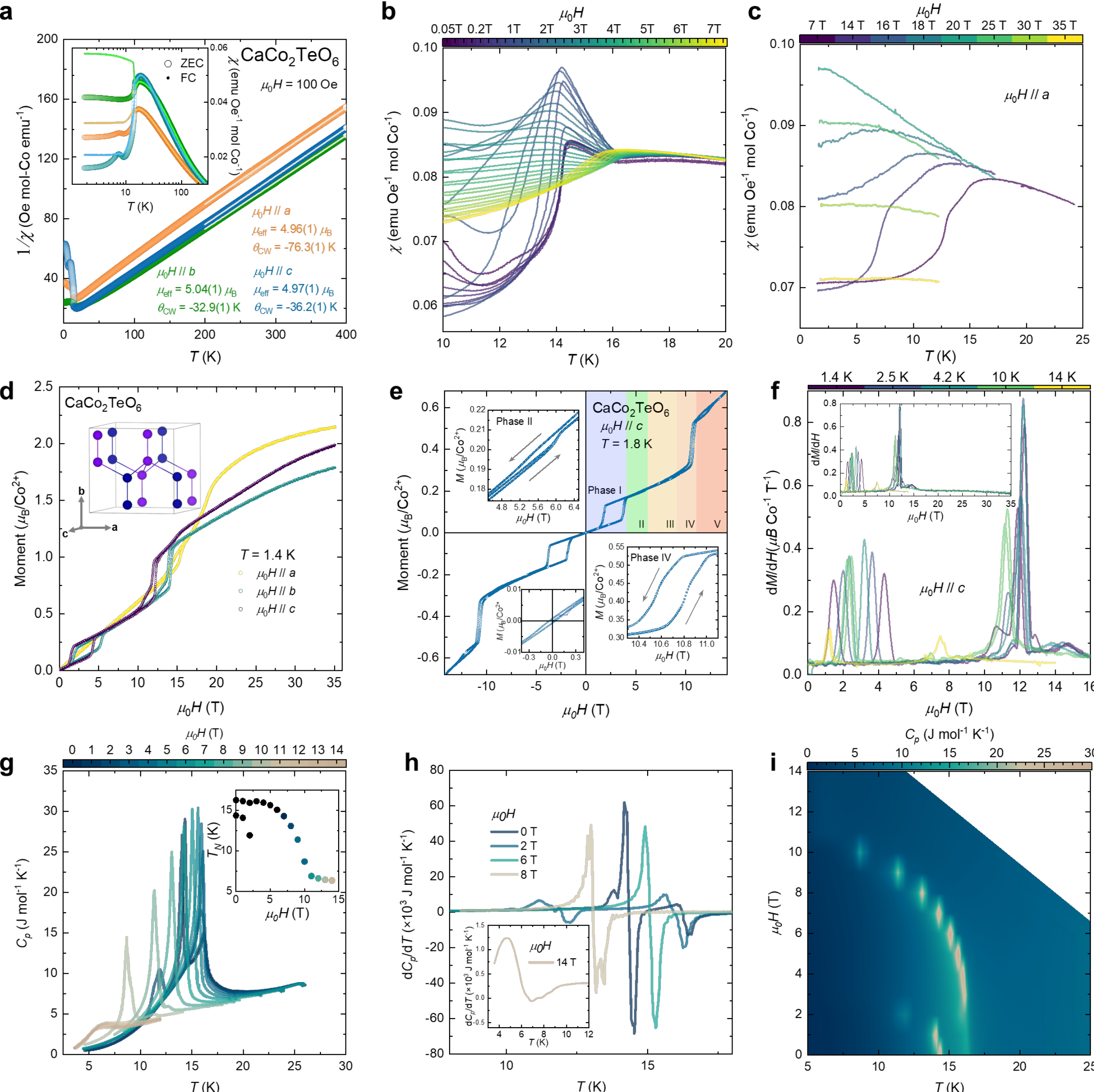

**Fig. 2.** (a) Inversed orientation dependent magnetic susceptibility vs. temperature with Curie-Weiss analysis at $\mu_0 H$ =100 Oe, showing the extracted moment and magnetic anisotropy. (b) Temperature-dependent magnetic susceptibility under various magnetic fields illustrating suppressed magnetic ordering. (c) Magnetic susceptibility vs. $T$ under high magnetic fields. (d) Orientation dependent magnetic moment as a function of applied field at $T$ = 1.4 K. (e) $M(H)$ curve at $T$ = 1.8 K showing field-induced tunability of magnetic phases. (f) First derivative of $M(H)$ at $\mu_0 H \parallel c$ and different temperatures. (g) Heat capacity revealing two anomalies that shift to lower temperature under increasing magnetic fields and evolve into a broad hump; inset shows the transition temperature as a function of applied field. (h) Derivative $dC_p/dT$ showing the evolution of magnetic transitions with fields. (i) Color map of heat capacity as a function of temperature and magnetic field, tracking the progression of magnetic transitions.

No frequency dependence is observed in the AC magnetization data under different fields (Fig. S7), indicating no spin-freezing transition. The isothermal magnetic entropy change is derived from the Maxwell relation (Equation 1):

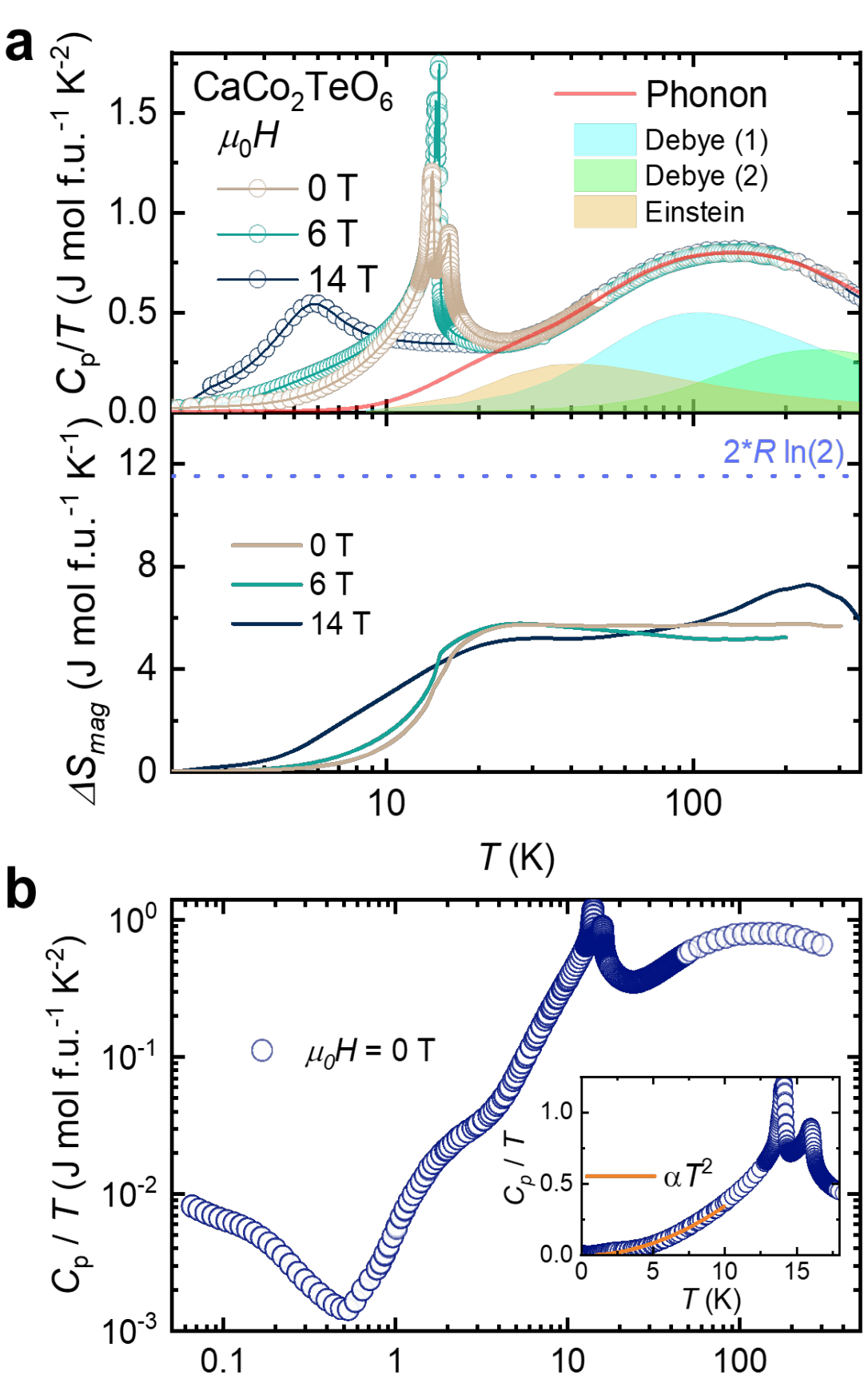

**Fig. 3.** (a) Molar heat capacity over temperature ($C_p/T$) vs. temperature for $CaCo_2TeO_6$ at $\mu_0H$ = 0 T and calculated phonon (red). The anomalies are consistent with the magnetic phase transitions of the material. Magnetic entropy change ($\triangle S_{mag}$) at different magnetic fields (solid lines) compared to the expected value of $J_{eff}$ = 1/2 spins (2*Rln2) (dash line). (b) Low-temperature heat capacity of $CaCo_2TeO_6$ showing the onset of the nuclear quadrupole Schottky Co at 0.06 K ≤ T ≤ 0.4 K, with the insert showing the characteristic $C_p = \alpha T^3$ behavior for gapless AFM magnon.

$$\left(\frac{dS}{dH}\right)_T = \left(\frac{dM}{dT}\right)_H \quad (1)$$

where $S$ is the entropy, $M$ is the magnetization, $T$ is the temperature, and $H$ is the magnetic field. The magnetic entropy can be extracted from Equation 2:

$$\Delta S_{mag}(H,T) = \int_0^H \left(\frac{dM}{dT}\right)_{H'} dH' \quad (2)$$

$\Delta S_{mag}$ $(H,T)$ map reveals a red region of positive entropy of approximately 1.5 J mol$^{-1}$ K$^{-1}$ at 15–16 K and above 6 T, indicative of the formation of topologically nontrivial phases (Fig. S5c).

The field-induced magnetic transitions are further confirmed by specific heat measurements at 0–14 T (Fig. 2g-i). At low field 0–4 T, two peaks are observed in $C_p(T)$, denoting the two magnetic phase transitions as seen in the magnetization data. As magnetic fields increase, these peaks shift to lower temperatures and become broader. At high field 4–14 T, these peaks evolve into one, which is eventually suppressed above 10 T. This vanishing indicates that the magnetic ordering is suppressed and the $O_h$-$Co^{2+}$ diamond system enters a quantum fluctuation state with short-range correlations at high fields. Similar field-driven quantum spin liquids or spin dynamics have been observed in honeycomb magnets α-$RuCl_3$, $BaCo_2(AsO_4)_2$, $Na_2Co_2TeO_6$, and $CaCo_2TeO_6$.[25,34-39] This similarity may stem from these honeycomb magnets and the $O_h$-$Co^{2+}$ diamond possessing bipartite lattices, which foster competing nearest-neighbor and next-nearest-neighbor interactions at a comparable energy scale (Fig. S8).[40,41] While some degree of perturbation of magnetic ordering has also been observed in $T_d$-$Co^{2+}$ diamond, the realization of the $J_{eff}$ = 1/2 ground state over a wide temperature range and field-induced quantum disorder is unique for $O_h$-$Co^{2+}$ $CaCo_2TeO_6$.

Fig. 3 presents the heat capacity at 0.1–300 K and 0 T, and phonon and magnetic entropy analysis. The magnetic entropy change $\Delta S_{mag}$ can be calculated from Equation 3:

$$\Delta S = \int_0^T \frac{C_v}{T} dT \quad (3)$$

where $C_v$ is the heat capacity at constant volume, which is approximated to be $C_p$ (heat capacity at constant pressure) for solids at low temperatures, and $T$ is the temperature. We constructed a phonon model that best describes the high-temperature heat capacity data by using two Debye modes and one Einstein model as follows (Equation 4):

$$\frac{C_p}{T} = \frac{C_{Debye(1)}}{T} + \frac{C_{Debye(2)}}{T} + \frac{C_{Einstein(1)}}{T} \quad (4)$$

$$C_{Debye} = 9NRs_D \left(\frac{T}{\theta_D}\right)^3 \mathrm{D}(\theta_D/T) \quad (5)$$

$$C_{Einstein} = 3NRs_E \frac{(\theta_E/T)^2 \exp(\theta_E/T)}{[exp(\theta_E/T) - 1]^2} \quad (6)$$

where $N$ represents the number of atoms, $R$ is the gas constant, $T$ is the sample temperature, $s_D$ is the number of oscillators of acoustic phonon, $\theta_D$ is the Debye temperature, $D(\theta_D/T)$ is the Debye function, $s_E$ is the number of oscillators of optical phonon, and $\theta_E$ is the Einstein temperature. The model parameters from the fitting are summarized in Table S5. Our assessment of the two-Debye one-Einstein phonon model is based on the resulting good fit and physical oscillator terms. It is rationalized by the three subunits: the phonon modes of (i) the magnetic sublattice, (ii) the nonmagnetic sublattice, and (iii) the counteraction site in the structure framework. The total number of oscillators is 9.7(5), close to the total number of 10 atoms in the formula unit for $CaCo_2TeO_6$. While our phonon model in part agrees with the previously reported two-Debye model,[25] we justify the presence of the Einstein model based on the characteristics $T_{max}$ observed in the $C_p/T^3$ vs. $T$ plot (Fig. S9). After subtracting the phonon contribution, the magnetic heat capacity $C_{mag}/T$ vs. $T$ is integrated to obtain the entropy change $\Delta S_{mag}$ associated with the magnetic transition. Fig. 3a shows that $\Delta S_{mag}$ reaches a maximum of 5.7(1) J mol f.u.$^{-1}$ K$^{-1}$ above $T_{N1,2}$. A material having two $Co^{2+}$ ions with $J_{eff} = 1/2$ should have a magnetic entropy change of 2*Rln(2) = 11.52 J mol f.u.$^{-1}$ K$^{-1}$. The observed $\Delta S_{mag}$ is only ~1/2 of the expected value for $J_{eff} = 1/2$ up to 300 K. The missing entropy can be attributed to two possible reasons: (i) the two-Debye one-Einstein phonon model overestimates the lattice contribution, and (ii) quantum fluctuations and short-range correlations are present in $CaCo_2TeO_6$. The field-induced magnetic entropy change $\Delta S_{mag}^H$ can be extracted by taking the difference between $\Delta S$ ($\mu_0 H$, $T$) and $\Delta S$ (0, $T$) (Fig. S5d-f). A positive $\Delta S_{mag}^H$ peak implies a high entropy magnetic state, whereas a negative peak denotes a long-range magnetic ordering. The evolution of positive and negative $\Delta S_{mag}^H$ peaks as a function of temperature at different magnetic fields confirms the rich magnetic properties of the material, consistent with the magnetization data. As the magnetic field increases, the long-range magnetic ordering (negative peak) is suppressed and transitions into a high-entropy magnetic state. This evolution indicates the field-induced tunability of the phase space. The heat capacity at 0.5–10 K below $T_N$ follows a $C_p = \alpha T^3$ behavior (Fig. 3b insert), which is expected for gapless antiferromagnetic magnons.[12] Fig. 3b shows an upturn in the low-temperature heat capacity at 0.06–0.4 K, which is too low in energy and magnitude to be assigned as a magnetic phase transition. It is most likely to be attributed to the nuclear quadrupole Schottky from Co, however, the maximum of such Schottky anomaly is expected to be around the temperature of $10^{-3}$ K.[42] While the observed Schottky tail does not supply sufficient information to be extracted for the nuclear quadrupole Schottky of Co, it proves the realization of quantum fluctuations down to $T$ = 0.06 K.

Fig. 4a-d presents powder neutron diffraction data that reveal strong temperature- and field-dependent behaviors, corroborating the magnetic phase diagram determined from heat capacity and magnetization. Specifically, the (110) magnetic Bragg peak is only present upon applied fields, and the intensity increases with magnetic field below $T_N$. In contrast, the intensity of the strongest (111) magnetic Bragg peak decreases with applied fields. These neutron results confirm an ordering temperature of 16.5(5) K at 0 T and 14.5(5) K at 5.5 T. As field is applied, new peaks emerge, possibly related to intermediate phases which are still under investigation. At 1.5 K and 0 T, the resulting magnetic peaks can be indexed by a single propagation vector, **k** = (0, 0, 0), with respect to the conventional unit cell. This corresponds to a magnetic structure in the *Pnma* (No. 62.441) magnetic space group ($\Gamma_1$ irreducible representation) and produces two distinct magnetic moments for each type of Co ion (Table S6-8). The resulting structure is a noncollinear AFM with the moment constrained within the *bc*-plane (Fig. 4f). Co(1) and Co(2) spins rotate ~56° and 64°, respectively, from the *b*-axis (Table S6). The refined magnetic moment of Co(1) is 2.3(1) $\mu_B$, ~55% of the static magnetic moment $\mu_{cal} = gS = 4.16\ \mu_B$ ($S$ = 3/2 and $g_{eff}$ = 2.77). The magnetic moment of Co(2) is determined to be 1.4(1) $\mu_B$, ~34% of the static moment and similar to that in the previous report.[25]

As temperature ramps from 1.5 K to 15 K, an intermediate magnetic phase is observed but difficult to determine with the powder data alone. To complement the powder diffraction data, we performed single-crystal neutron diffraction. Fig. 4e shows the 0KL plot measured at 14.5 K and 20 K. Satellite magnetic peaks occur at 14.5 K and can be indexed with a propagation vector **k** = (0, 0.2744(2), 0). The magnetic structure of the intermediate phase is shown in Fig. 4g and Fig. S11. The spin rotates 98.8(1)° both clockwise and counterclockwise in the *bc*-plane from the neighboring unit cell along the *b*-axis. The magnetic moments for Co(1) and Co(2) at 14.5 K are extracted to be 2.388(1) $\mu_B$, and 2.165(1) $\mu_B$, only ~57% and 52% of the static moment, respectively (Table S6, 9). With this insight from the single-crystal neutron data, the propagation vector is extracted from the powder diffraction data at 15 K to be **k** = (0, 0.25(5), 0)—close to a commensurate wavevector of 1/4. This realization from the powder data enabled us to determine the magnetic structure with a comparable propagation vector and the Co(1) and Co(2) magnetic moments of 1.65(1) $\mu_B$, and 2.03(1) $\mu_B$, ~40% and 50% of the static moment, respectively (Table S6, S10, Fig. S11). The small variation of the propagation vector between the single-crystal and powder neutron data might be attributed to a slight difference in temperature at the sample position. The appreciable reduction in the static moment on both the Co(1) and Co(2) sites at different temperatures proves the presence of considerable quantum fluctuations, as revealed in the missing entropy. Similar behaviors have been observed in $S$ = 3/2 $ZnCr_2Se_4$ pyrochlore [43] and $S$ = ½ $CuAl_2O_4$ diamond.[44] Nevertheless, the realization of the $J_{eff}$ = 1/2 ground state over a wide temperature window with strong quantum fluctuations differentiates $CaCo_2TeO_6$ from other frustrated magnets.

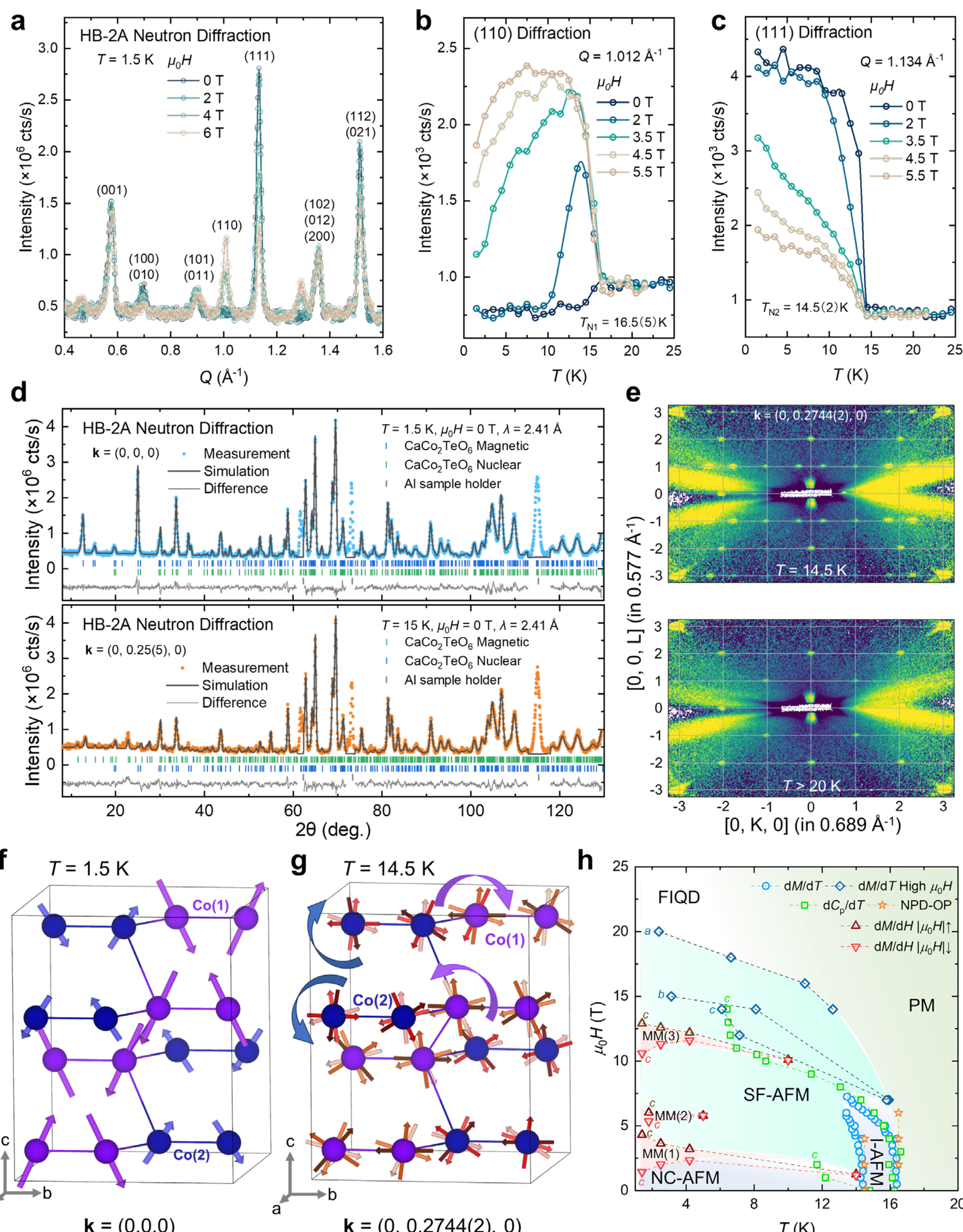

**Fig. 4.** (a) Powder neutron diffraction of $CaCo_2TeO_6$ at a wavelength of 2.41 Å under different fields at $T$ = 1.5 K. (b) Evolution of the intensity of the magnetic Bragg peaks (110) at 1.5 K ≤ T ≤ 25 K under different fields, showing the Neel temperature of $T_{N1}$ = 16.5(5) K (c) The intensity of the most intense magnetic Bragg peak (111) at different fields showing the Neel temperature of $T_{N2}$ = 14.3(2) K. (d) Magnetic structure refinement on the powder data at 1.5 K and 14.5 K. (e) Single crystal neutron diffraction pattern at 15 K and $T$ > 20, the additional satellite peaks occur at 15 K below $T_{N1}$ indicate the propagation vector of (0, 0.2744, 0). (f) Refined magnetic structure at 1.5 K with the propagation vector **k** = (0, 0, 0). (g) Refined incommensurate magnetic structure at 14.5 K from single crystal neutron diffraction. (h) Magnetic phase diagram. Labeled lines show field orientation of orientation-dependent data on a single crystal, with labels indicating the magnetic field direction; unlabeled lines represent powder data.

Fig. 4h presents a magnetic phase diagram mapped out based on the combination of physical properties and neutron measurements. The phase boundaries and the order of the phase transitions are informed by magnetic entropy and heat curves (Fig. S13, S14). Below 10.5 T, the system displays a first-order transition from a paramagnetic (PM) state into an incommensurate antiferromagnetic (I-AFM) phase at 14.5−16.5 K, before settling into a noncolinear AFM (NC-AFM) ground state. Upon increasing field, the NC-AFM phase exhibits metamagnetic transitions (MM), as evidenced by pronounced hysteresis (Fig. 2e-f, S6), and then evolves into a high-entropy spin-flop AFM (SF-AFM) phase (Fig. S5e-f). At 10.5 T ≤ $\mu_0 H$ ≤ 14 T, heat capacity and magnetization measurements reveal a broad anomaly near 6.5 K (Fig. 2, S7, S13–S14), reminiscent of a second-order transition; however, no direct evidence of symmetry breaking has been observed. At higher fields, magnetic order is progressively suppressed, and the last observable AFM transition occurs at $T_N$ = 2.5(5) K and $\mu_0 H$ = 20 T (Fig. 2c), revealing that the system ultimately bypasses long-range order and enters a field-induced quantum disordered (FIQD) state.

Fig. 5a-b highlights the comparison of chemical bonding and electronic instability of $T_d$-Co $CoRh_2O_4$ vs $O_h$-Co $CaCo_2TeO_6$ through our assessment of their integrated projected crystal orbital Hamilton population (ICOHP) values (Fig. S16-S21). $CaCo_2TeO_6$ displays more considerable antibonding around the Fermi level and thus larger electronic instability (7.79%) compared to $CoRh_2O_4$ (3.01%), suggesting that pathways to exotic states of matter in $CaCo_2TeO_6$ are more accessible than in conventional diamond magnets. We extracted $J$-coupling constants and validated multicenter Co-O-Co interactions through crystal orbital bond index (COBI) analysis (Fig. S22-24).[45] The integrated COBI (ICOBI) implies that the interactions between the same Co types Co(1)-O-Co(1) ($J_{1\text{-}2}$) and Co(2)-O-Co(2) ($J_{3\text{-}4}$) surpass those between different Co types Co(1)-O-Co(2) ($J_{5.6}$) (Fig. S24e). To account for the direct Co-Co interactions, we calculated the COHP between the Co atoms within a tetrahedron. The ICOHP confirms that the Co(1)-Co(1) interaction is stronger than Co(1)-Co(2) (Fig. S24f).

To quantify the Co-Co magnetic exchange interactions (Fig. 5c-d), we applied the Green's function method by using the Wannier formalism and the Heisenberg model.[46] The nearest-neighbor magnetic exchange interactions $J_{1\text{-}4}$, facilitated by edge-sharing $O_h$-[$CoO_6$], are identified to be $J_1$ = -49 K (AFM), $J_2$ = 43 K (FM), $J_3$ = -53 K (AFM) and $J_4$ =43 K (FM). The magnitude of the calculated $J$-constants aligns well with the extracted Curie-Weiss temperature, and the sign is in excellent agreement with the magnetic ground state determined from the neutron experiments. The next-nearest-neighbor magnetic exchange interactions, mediated by corner-sharing $O_h$-[$CoO_6$], are estimated to be $J_5$ = -27 K (AFM) and $J_6$ = -23 K (AFM), in the same order of magnitude as the nearest-neighbor interactions $J_{1\text{-}4}$. This substantiates three-dimensional Heisenberg magnetic interactions in $CaCo_2TeO_6$—fundamentally different from the 2-D honeycomb model used in the recent report.[25] The strong competing Heisenberg AFM-FM nearest-neighbor and next-nearest-neighbor exchange interactions at a similar energy scale are essential for the putative quantum nature of the spin fluctuations.[47-49] The second-nearest-neighbor exchange interactions $J_{7\text{-}8}$ are about 1/8 of $J_{1\text{-}6}$, whereas the third-nearest-neighbor interactions $J_{9\text{-}10}$ are negligible. This is similar to what has been observed in conventional diamond magnets.[16,50] Nevertheless, the strong competing AFM-FM coupling constants and unique exchange pathways in the $O_h$-$Co^{2+}$ diamond magnet offer a new avenue to realize the $J_{eff}$ = 1/2 ground state with strong quantum fluctuations and a field-induced quantum disordered state.

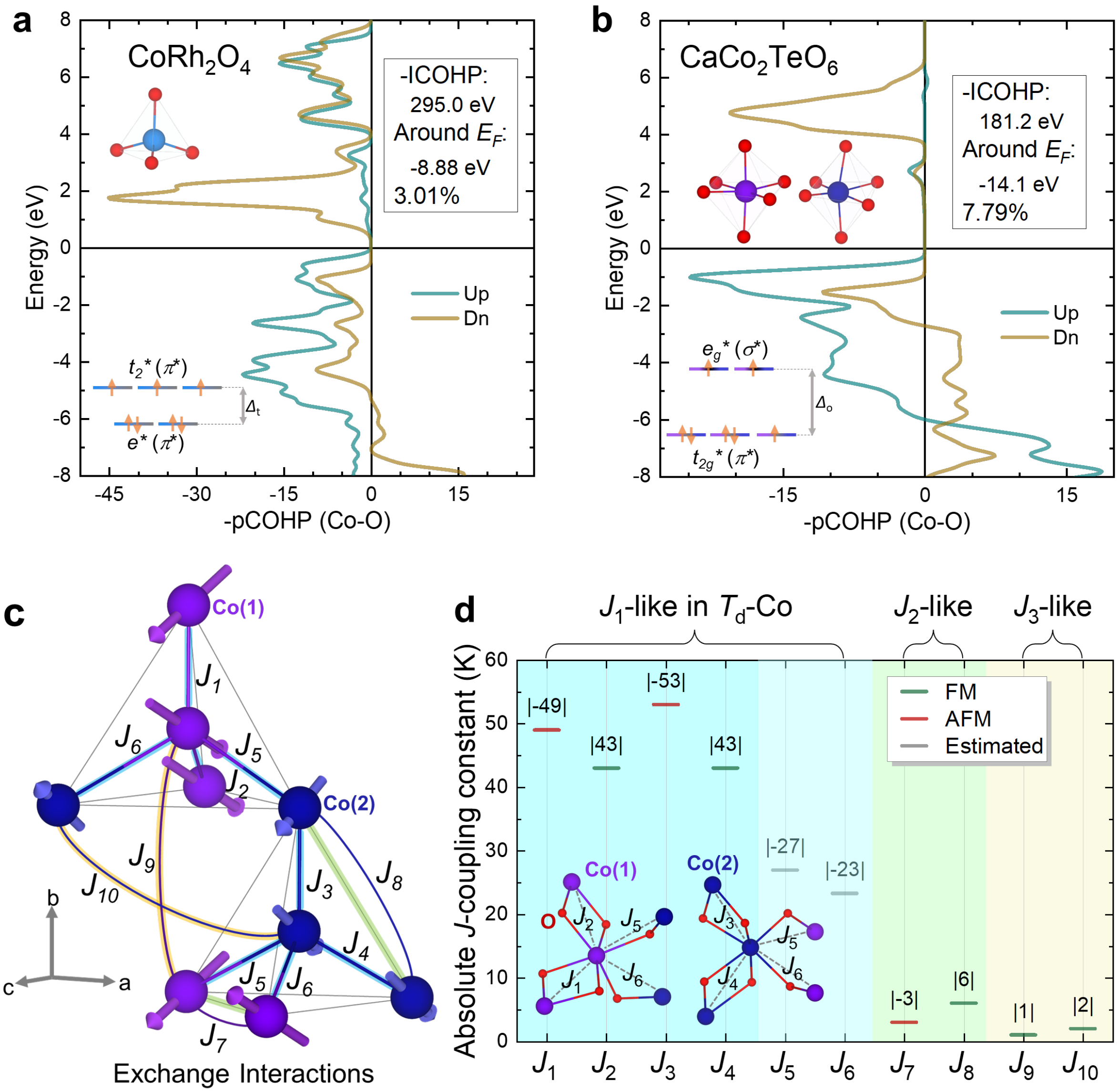

**Fig. 5.** Projected crystal orbital Hamilton population (pCOHP) for (a) $CaCo_2TeO_6$ and (b) $CoRh_2O_4$with their integrated value (ICOHP) indicating antibonding states near $E_F$ and enhanced electronic instability in Oh-$Co^{2+}$ around $E_F$. (c) *J*-coupling interactions in $CaCo_2TeO_6$ at the magnetic ground state. (d) Calculated *J*-coupling constants, showing strong competing AFM and FM interactions at comparable energy scales.

## Conclusions

The results enable a new pathway for reimagining the modulation of quantum functions and spin dynamics in geometrically frustrated magnets through ligand-field engineering. We present $CaCo_2TeO_6$, a new diamond magnet where two distinct $Co^{2+}$ ions experience an $O_h$ ligand field—an untapped feature in diamond-lattice systems. This adds a new dimension beyond conventional spinel diamond magnets with $T_d$-site magnetic ions. The material offers new frontiers in field-tunable spin physics, strong competing AFM-FM exchange interactions, pronounced quantum fluctuations down to 0.06 K, and field-induced quantum disorder. At high fields, two AFM magnetic orders are suppressed to lower temperatures and then evolve into one transition, which eventually vanishes above 20 T. Although the net magnetic interactions are dominated by AFM exchange, the double hysteresis loops in $M(H)$ reveal competing FM correlations. The material displays a noncolinear AFM magnetic ground state at base temperature and an incommensurate AFM intermediate state at 15 K. The significant reduction in the static magnetic moment on both Co sites further confirms strong quantum fluctuations. High-field measurements reveal the complete suppression of long-range magnetic order. The absence of magnetic order above 20 T, combined with persistent quantum fluctuations, suggests the emergence of a field-induced quantum disordered state, potentially a quantum spin liquid. We construct a magnetic phase diagram based on these experimental results to guide the tantalizing phase space of $CaCo_2TeO_6$. Complementary DFT calculations highlight covalent Co-O bond characters and increased electronic instability, in addition to strong competing Heisenberg nearest-neighbor and next-nearest-neighbor magnetic exchange interactions at comparable energy scales. These enhanced competing interactions arise from the new coupling pathways—edge- and corner-sharing [$CoO_6$] octahedra—in the $O_h$-$Co^{2+}$ diamond magnet. Our work demonstrates a simple, innovative approach to using ligand fields for atomically modifying the energy landscape of many-body electronic states and the competing Heisenberg exchange interactions. The realization of the novel $J_{eff}$ = 1/2 ground state at elevated temperatures, along with strong quantum fluctuations and a putative field-induced quantum spin liquid state, underscores a profound manifestation of ligand fields and spin-orbit coupling. This study represents a rare example of frustrated magnets that offers new insights both experimentally and theoretically for potential quantum functions.

## References and Notes

**Acknowledgments:** This work was supported by the Arnold and Mabel Beckman Foundation grant 2023 BYI and the Camille and Henry Dreyfus Foundation grant 2025 Camille Dreyfus Teacher-Scholar award TC-25-071. TTT acknowledges the U.S. National Science Foundation grant NSF-DMR-CAREER-2338014 and grant NSF-OIA-2227933. XH and TTT thank Dr. C. McMillen, Dr. R. Sachdeva, and Dr. Matthew Powell for their assistance in X-ray diffraction, TGA, and some physical property measurements. The work by YC and RB at the University of California, Berkeley and Lawrence Berkeley National Laboratory was funded by the U.S. DOE, Office of Science, Office of Basic Energy Sciences, Materials Sciences and Engineering Division under Contract No. DE-AC02-05CH11231 (Quantum Materials Program KC2202). LPC and JA acknowledge the EPiQS Initiative of the Gordon and Betty Moore Foundation through grant no. GBMF9067. The research performed at the Gdańsk Tech was supported by the National Science Centre (Poland) under SONATA-15 grant (UMO-2019/35/D/ST5/03769). A portion of this research used resources at the High Flux Isotope Reactor and Spallation Neutron Source, a DOE Office of Science User Facility operated by the Oak Ridge National Laboratory. The beam time was allocated to HB-2A POWDER on proposal number IPTS-31480.1 and CORELLI on proposal number 34873.1. T.K. and H.C. acknowledge the support by the U.S. Department of Energy, Office of Science, Office of Basic Energy Sciences Biopreparedness Research Virtual Environment program. T.K. and H.C. thank Zachary Morgan for his assistance during the data reduction. Use of the Advanced Photon Source at the Argonne National Laboratory was supported by the U.S. Department of Energy, Office of Science, Office of Basic Energy Sciences, under Contract No. DE-AC02-06CH11357. A portion of this work was performed at the National High Magnetic Field Laboratory, which is supported by National Science Foundation Cooperative Agreement No. DMR-2128556 and the State of Florida. This manuscript was authored by UT-Batelle, LLC, under contract DE-AC05-00OR22725 with the US Department of Energy (DOE). The US government retained and the publisher, by accepting the article for publication, acknowledged that the US government retained a nonexclusive, paid-up, irrevocable, worldwide license to publish or reproduce the published form of this manuscript, or allowed others to do so, for US government purposes. DOE would provide public access to these results of federally sponsored research in accordance with the DOE Public Access Plan (http://energy.gov/downloads/doe-public-access-plan).

**Author contributions:**

Conceptualization: XH, TTT

Methodology: XH, LPC, BD, MJW, MS, XZ, YC, DY, TK, HC, ESC

Investigation: XH, LPC, BD, MJW, MS, XZ, YC, DY, TK, HC, ESC

Visualization: XH, LPC, DY, TK, HC

Funding acquisition: MJW, AK, RB, JA, SC, DY, HC, TTT

Project administration: XH, TTT

Supervision: MJW, AK, RB, JA, SC, DY, HC, TTT

Writing – original draft: XH, LPC, DY, HC

Writing – review & editing: XH, LPC, BD, MJW, MS, XZ, YC, AK, RB, JA, SC, DY, TK, HC, ESC, TTT

**Competing interests:** Authors declare that they have no competing interests.

**Data and materials availability:** All data are available in the main text and the supplementary materials.

Supplementary Materials for

# $J_{eff}$ = ½ Diamond Magnet $CaCo_2TeO_6$: Reimagining Frontiers of Spin Liquids and Quantum Functions

Xudong Huai,[1] Luke Pritchard Cairns,[2] Bridget Delles,[1] Michał J. Winiarski,[3] Maurice Sorolla II,[4] Xinshu Zhang,[5] Youzhe Chen,[2] Stuart Calder,[6] Eun Sang Choi,[7] Tatenda Kanyowa,[8] Anshul Kogar, [5] Huibo Cao,[6,8] Danielle Yahne,[6] Robert Birgeneau,[2] James Analytis,[2] Thao T. Tran*[1]

*thao@clemson.edu

**The PDF file includes:**

## Experimental Section

### Reagents

The starting materials are $CaCO_3$ (Acros organics, 99%), $CoCO_3$ (Alfa Aesar, 99.5%), $Te(OH)_6$ (Tokyo Chemical Industry, 99.0%), and $TeO_2$ (Alfa Aesar, 99%).

### Synthesis

A small amount of crystal is first prepared by flux growth. $CaCO_3$, $CoCO_3$, and $Te(OH)_6$ powder (molar ratio 2:2:5) was ground and pressed into a pellet. The pellet is then sent to a box furnace, heated at 850°C, followed by slow cooling. With the crystal, we acquired the melting point (625°C) and decomposition point (800°C) of $CaCo_2TeO_6$ by conducting TGA-DSC measurements.

With the melting point confirmed, the polycrystalline sample was prepared by a solid-state reaction. A stoichiometric mixture of $CaCO_3$, $CoCO_3$, and $Te(OH)_6$ powder was ground and pressed into a pellet and then placed in a box furnace with the temperature set at 625 °C for eight days with multiple grinding and pelletizing. Powder XRD is then performed on this pinkish-purple powder, and the structure of the polycrystalline sample is confirmed by Rietveld refinement.

High-quality crystal samples were then synthesized by flux growth method by heating polycrystalline $CaCo_2TeO_6$ and $TeO_2$ as flux (molar ratio 2:5) at 900°C for 40 h followed by slow cooling. Block-shaped dark blue crystals can be separated from the transparent flux. The reason why $Te(OH)_6$ is not selected as self-flux is because its molten phase does not have enough mobility to dissolve $CaCo_2TeO_6$.

### Single-Crystal X-ray Diffraction

Single-crystal XRD experiments were performed on $CaCo_2TeO_6$ using a Bruker D8 Venture diffractometer with Mo Kα radiation ($\lambda = 0.71073$ Å) and a Photon 100 detector at $T = 300$ K. Data processing (SAINT) and scaling (SADABS) were performed using the APEX3 software. The structure was solved by the intrinsic phasing 2 method (SHELXT) and refined by full-matrix least-squares techniques on F2 (SHELXL) using the SHELXTL software suite. All atoms were refined anisotropically.

### Synchrotron X-ray Diffraction

A synchrotron XRD pattern of $CaCo_2TeO_6$ was collected using the 11-BM beamline at Advanced Photon Source, Argonne National Laboratory. Data were collected using a well-grounded crystal of $CaCo_2TeO_6$ at $T = 295$ K and $\lambda = 0.45789$ Å. Rietveld refinement of the XRD pattern was performed using TOPAS Academic V6. [1, 2] Vesta software was used for crystal structure visualization.

### Vis-NIR Spectroscopy

$CaCo_2TeO_6$ powder is ground with PTFE and pressed into a pellet. The reflectance is then measured from 2500 – 480 nm (0.5 – 2.5 eV) by Agilent Cary 7000 UV-Vis-NIR spectrometer.

### Second Harmonic Generation (SHG)

We performed second harmonic generation spectroscopy on $CaCo_2TeO_6$ in transmission configuration. The incident light pulse was generated by an optical parametric amplifier with tunable wavelength 950-1030 nm (roughly 1.2-1.3eV). The incident light was focused normally on the sample with a 100 μm spot size, and the fluence was kept below 1 mJ/cm$^2$ to minimize heating. The light interacting with the sample generated a second harmonic radiation. The detection of the second harmonic light was conducted with a commercial photo-multiplier tube. The sample temperature can be varied from 4 K to 300K with a standard optical cryostat.

**Magnetization and Specific Heat**

DC magnetization measurements on $CaCo_2TeO_6$ powder were performed with the vibrating sample magnetometer (VSM) option of Quantum Design Physical Properties Measurement System (PPMS). AC magnetization was measured using the ACMS option. Magnetic susceptibility was approximated as magnetization divided by the applied magnetic field: $\chi \approx M/H$. Specific heat measurements were performed using a Quantum Design PPMS Dynacool with $^3$He insert for T > 1 K and a home-built dilution refrigerator setup for T < 1 K. The data was acquired using the relaxation time method across the full temperature range.

**High-field Magnetization**

Orientation-dependent high-field magnetization measurements of $CaCo_2TeO_6$ were performed on a 13.73 mg single crystal using a conventional VSM coupled with a water-cooled resistive magnet in Cell 8 of the DC Field Facility at the National High Magnetic Field Laboratory (NHMFL) in Tallahassee, Florida. The VSM was calibrated against a nickel-standard sphere, and the sample was mounted on a brass holder using 3.5 mm quartz rods and GE-7031 varnish. To further calibrate the temperature and field, the resulting magnetization data were normalized by comparing them with a DC magnetization curve obtained via a VSM in a 16 T PPMS (Quantum Design) at NHMFL.

**Powder Neutron Diffraction**

A 5 g powder sample of $CaCo_2TeO_6$ was pressed into pellets and loaded into an Al can, then backfilled with 1 atm He. The sample was then loaded into a cryomagnet on the HB-2A powder diffractometer at ORNL. Measurements were performed down to 1.6 K and up to 6 T magnetic field with the 2.41 Angstrom (Ge113) incident wavelength in the open-open-21' collimator setting. Magnetic structure determination was done using the FullProf[3] software suite, utilizing the Bilbao Crystallographic Server for the magnetic space group formalism, and SARAh Representational Analysis software for the irreducible representation formalism.

**Single-crystal Neutron Diffraction**

Single-crystal neutron diffraction measurements were conducted at CORELLI at the Spallation Neutron Source (SNS) at Oak Ridge National Laboratory. A single crystal (~3 mg) was mounted in a Closed-Cycle Refrigerator (CCR) for low-temperature measurements. Diffraction data were collected by rotating the crystal around the instrument's vertical axis. Measurements were taken during cooling (T > 17 K) and at T = 15 K. The difference between these datasets can be used to identify the ordered magnetic phase at 15 K. The well-indexed reciprocal space directly reveals the propagation vector of the magnetic order, which is difficult to determine from powder diffraction data. Structural refinement was performed using the FullProf Suite [3].

**Density Functional Theory Calculation**

Spin-polarized electronic structure calculations for $CaCo_2TeO_6$ were performed using a full-potential linearized augmented plane wave (FP-LAPW) method as implemented in the WIEN2k code[4]. The exchange-correlation potentials were treated within the density functional theory (DFT) using the Perdew−Burke−Ernzerhof generalized gradient approximation (PBE-GGA).[5] The self-consistencies were performed using 720 k-points ($10 \times 9 \times 8$ mesh) in the irreducible Brillouin zone. To correct for electron over-delocalization, a 9 eV, Hubbard U correction was applied for Co-*d* bands. The muffin-tin radius values 1.13, 1.22, 0.97, and 0.83 Å were used for Ca, Co, Te and O, respectively.

Pseudopotential band structure and polarized density of sates were calculated for $CaCo_2TeO_6$ and $CoRh_2O_4$ using the pw.x program in the Quantum Espresso (QE) software package,[6] with the Generalized Gradient Approximation of the exchange-correlation potential and Hubbard U correction (GGA+U) for Co-*d* orbital (9 eV) of the exchange-correlation potential with the PBEsol parametrization. Projector-augmented wave (PAW) potentials for Rh, Ca, Co, Te and O were taken from the PSlibrary v.1.0.0 set[7]. The applied k-mesh for $CaCo_2TeO_6$ is the same as in WIEN2k calculations, and a $8 \times 8 \times 8$ mesh was used for $CoRh_2O_4$. The kinetic energy cutoff for charge density and wavefunctions was set to 40 eV and 450 eV for $CaCo_2TeO_6$ (47 eV and 480 eV for $CoRh_2O_4$, respectively). The PAW wavefunctions calculated from the QE package are projected into a linear combination of atomic orbitals (LCAO) based representation by means of the Local Orbital Basis Suite Towards Electronic-Structure Reconstruction (LOBSTER) program[8, 9] to extract the projected crystal orbital Hamilton populations (-pCOHP ). [10, 11]

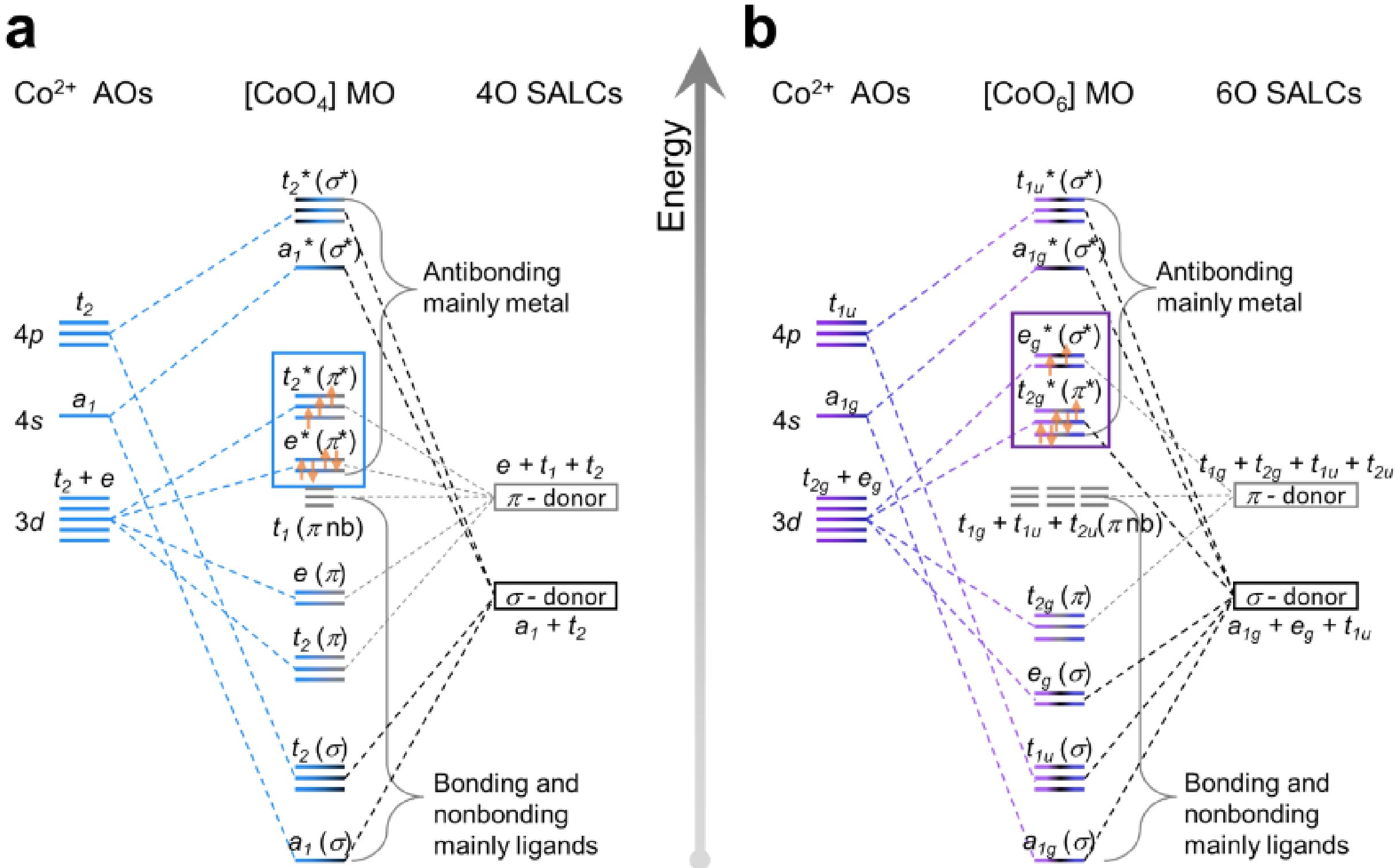

**Figure S1.** Molecular orbital interaction diagram for (a) $T_d$-[$CoO_4$] and (b) $O_h$-[$CoO_6$].

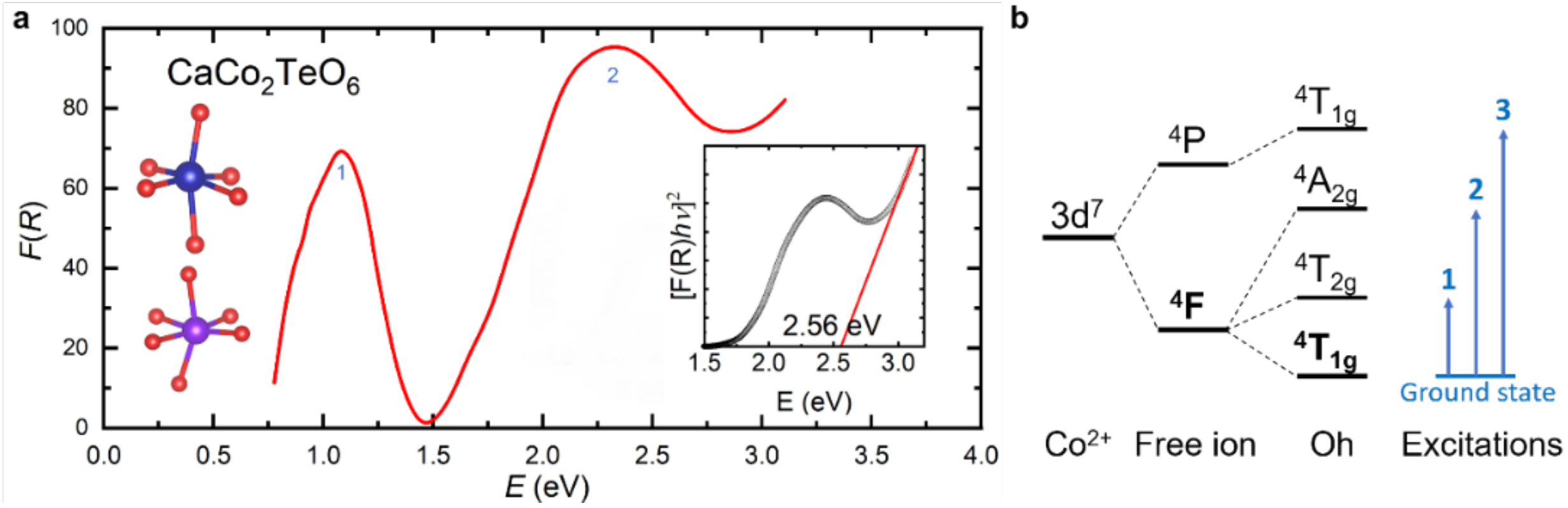

**Figure S2.** UV-Vis-NIR spectroscopy and the optical gap.

The obtained spectral profile (a) shows that the environments of $Co^{2+}$ ions closely resemble a near-ideal octahedral geometry (b). The optically determined band gap correlates well with the observed blue color of the crystal. Further insight is provided by the crystal field splitting diagram presented in Figure 1d, which elucidates the corresponding absorption (F(R)) spectrum observed.

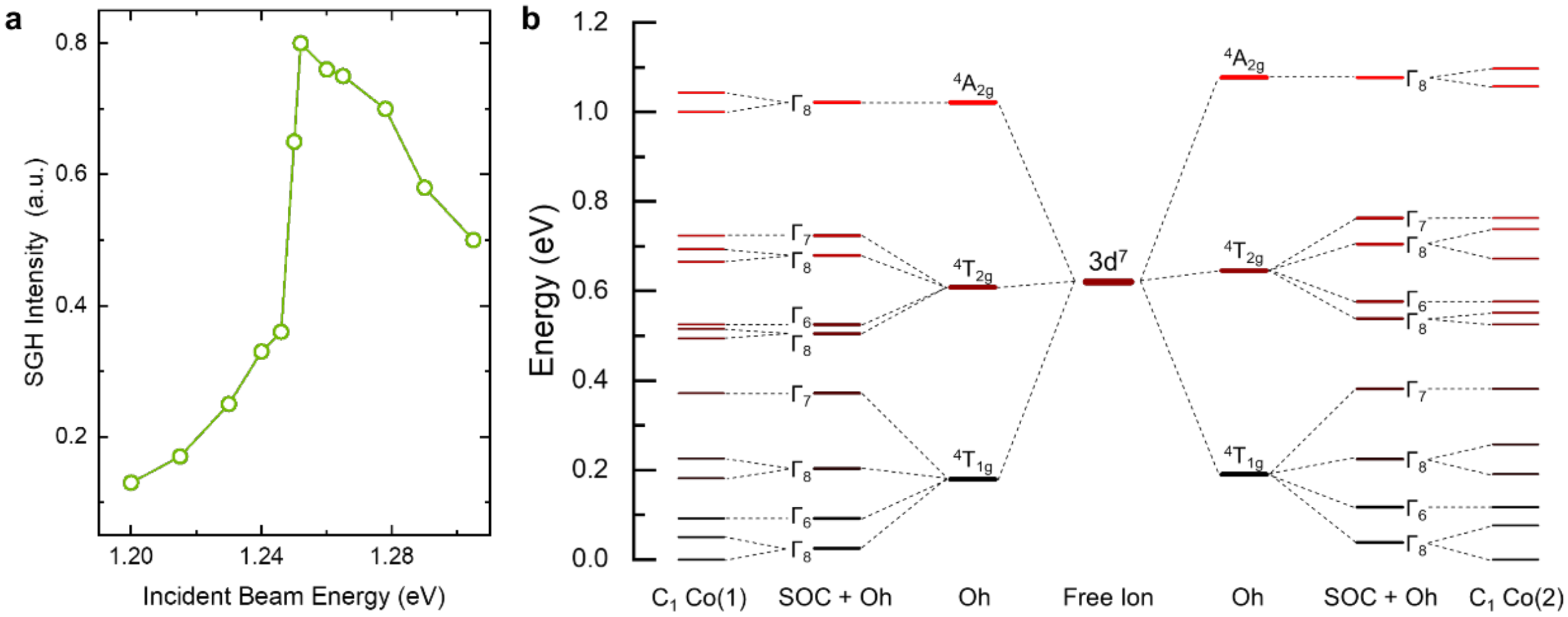

**Figure S3.** SHG spectrum and calculated crystal orbital splitting diagram with spin orbital coupling.

As the incident light resonates with the crystalline effect field (CEF) level of the materials (calculated from a point charge model using PyCrystalField), the SHG signal is dramatically enhanced compared with non-resonant SHG. Therefore, we can use SHG spectroscopy to precisely determine the CEF level of $CaCo_2TeO_6$. Although this material has inversion symmetry, the electronic dipole SHG is forbidden, the magnetic dipole SHG is allowed. We observed an intense peak around 1.25eV within the energy range in the SHG spectrum, which corresponds to the spin-allowed transition from the ${}^4T_{2g}$ to ${}^4T_{1g}$ state. SHG results are consistent with the absorption measurements and provide precise measurements of the CEF energy levels. The energy mismatch between calculated and measured values could arise from the covalency of the Co-O bond that is ignored by the point charge model. We didn't observe significant change in SHG intensity across the magnetic transition temperature ~16 K (4 - 100 K), indicating the origin of the SHG signal was from magnetic dipole or electronic quadrupole.

The calculated dipole moment vector (in arbitrary units):

$$[\ 1.28 \times 10^{-12},\ 7.39 \times 10^{-12},\ -3.82 \times 10^{-12}]$$

Dipole moment magnitude (in arbitrary units):

$$8.42 \times 10^{-12}$$

The calculated quadrupole moment tensor (in arbitrary units):

$$\begin{bmatrix} 1.35 \times 10^{3}, & 1.91 \times 10^{2}, & -4.91 \times 10^{3} \\ 1.91 \times 10^{2}, & 6.16 \times 10^{3}, & 1.76 \times 10^{3} \\ -4.91 \times 10^{3}, & 1.76 \times 10^{3}, & 7.51 \times 10^{3} \end{bmatrix}$$

Principal components of the quadrupole moment tensor:

$$[-6.56 \times 10^{3},\ -3.28 \times 10^{3},\ 9.84 \times 10^{3}]$$

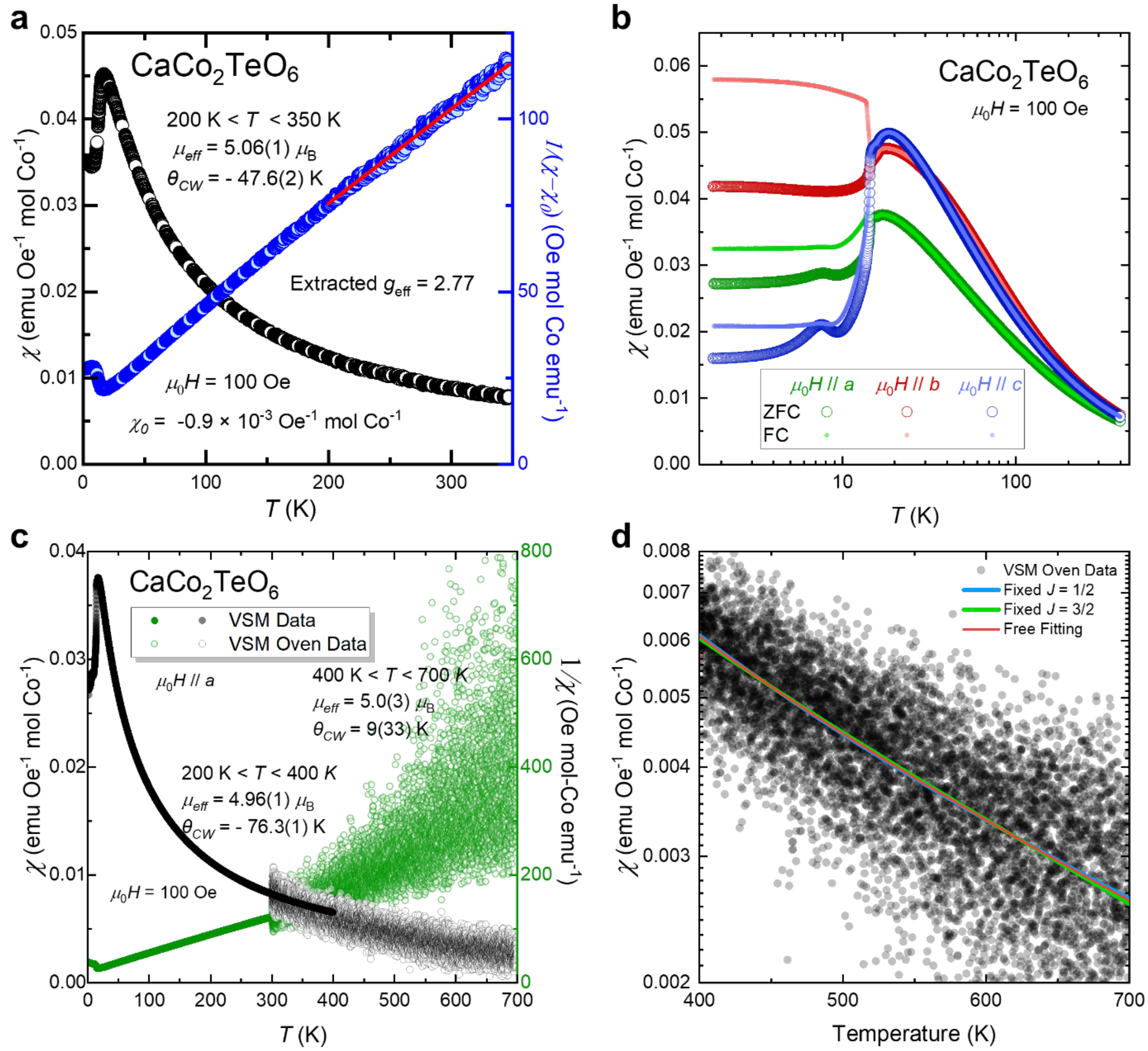

**Figure S4.** (a) Magnetic susceptibility from powder. (b) Orientation-dependent magnetic susceptibility as a function of temperature (c) High-temperature magnetic susceptibility. (d) Fittings on high-temperature magnetic susceptibility.

Fitted result from Fig. S4d:

| Fitting Type | $\theta_{CW}$ | $\mu_{eff}$ | $\chi_0$ | $R^2$ |
|---|---|---|---|---|
| Fixed $J = 1/2$ | 36(2) | 4.47 (Fixed, $S = 3/2$, $L = 1$) | -0.00161(3) | 0.61179 |
| Fixed $J = 3/2$ | -58(3) | 5.67 (Fixed, $S = 3/2$, $L = 2$) | -0.00273(3) | 0.61165 |
| Free Fitting | 9(33) | 5.01(3) | -0.00194(4) | 0.61178 |

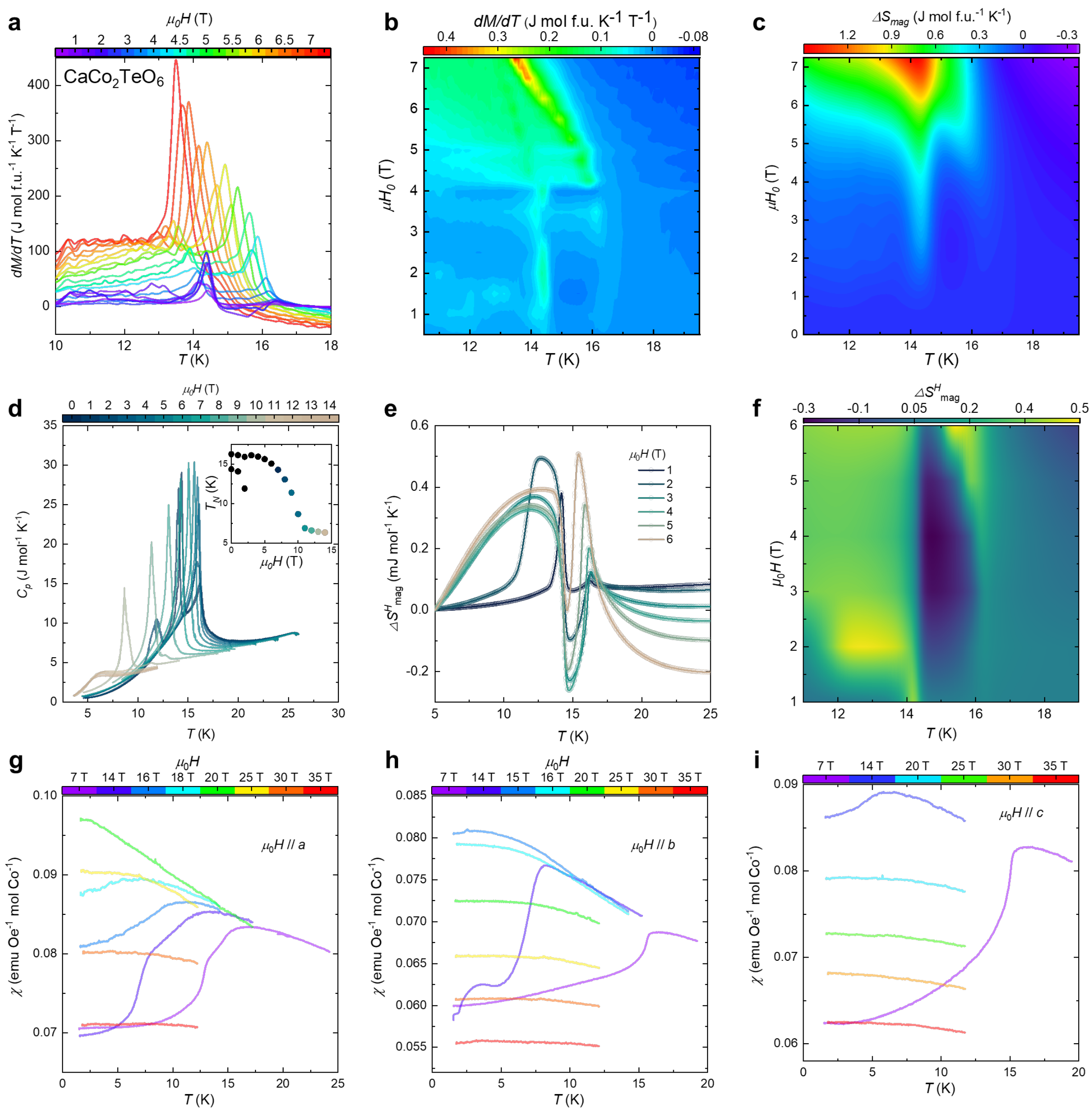

**Figure S5.** (a) First derivative of magnetization with respect to temperature under magnetic field. (b) A map of d$M$/d$T$ = d$S$/d$H$ showing two magnetic transitions. (c) A magnetic entropy map. (d) Field dependent heat capacity. (e, f) magnetic entropy change with respect to field extracted from heat capacity. (g-i) Temperature dependent magnetic susceptibility along different directions under high magnetic field.

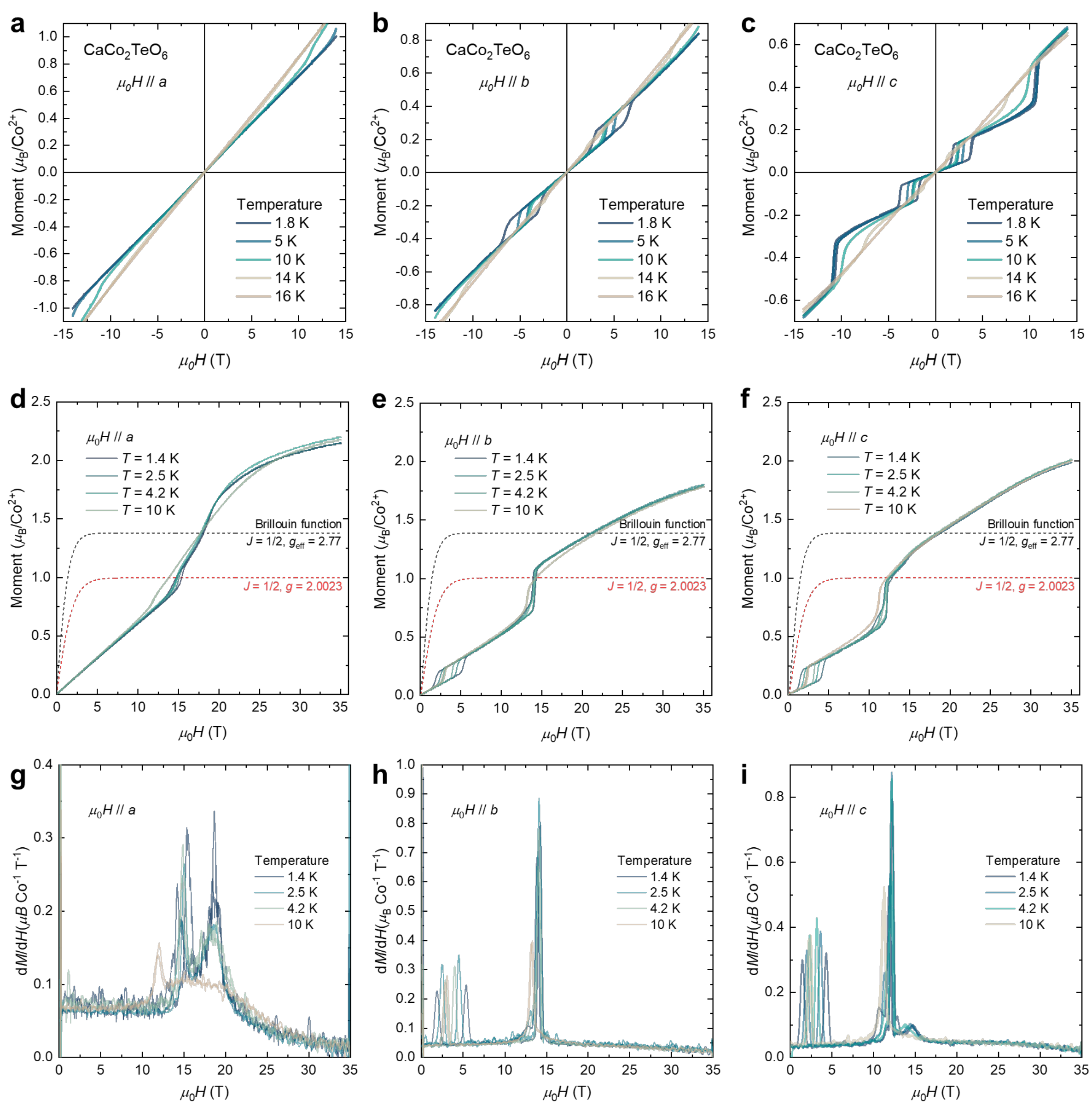

**Figure S6.** Orientation-dependent magnetization under magnetic field up to 35 T.

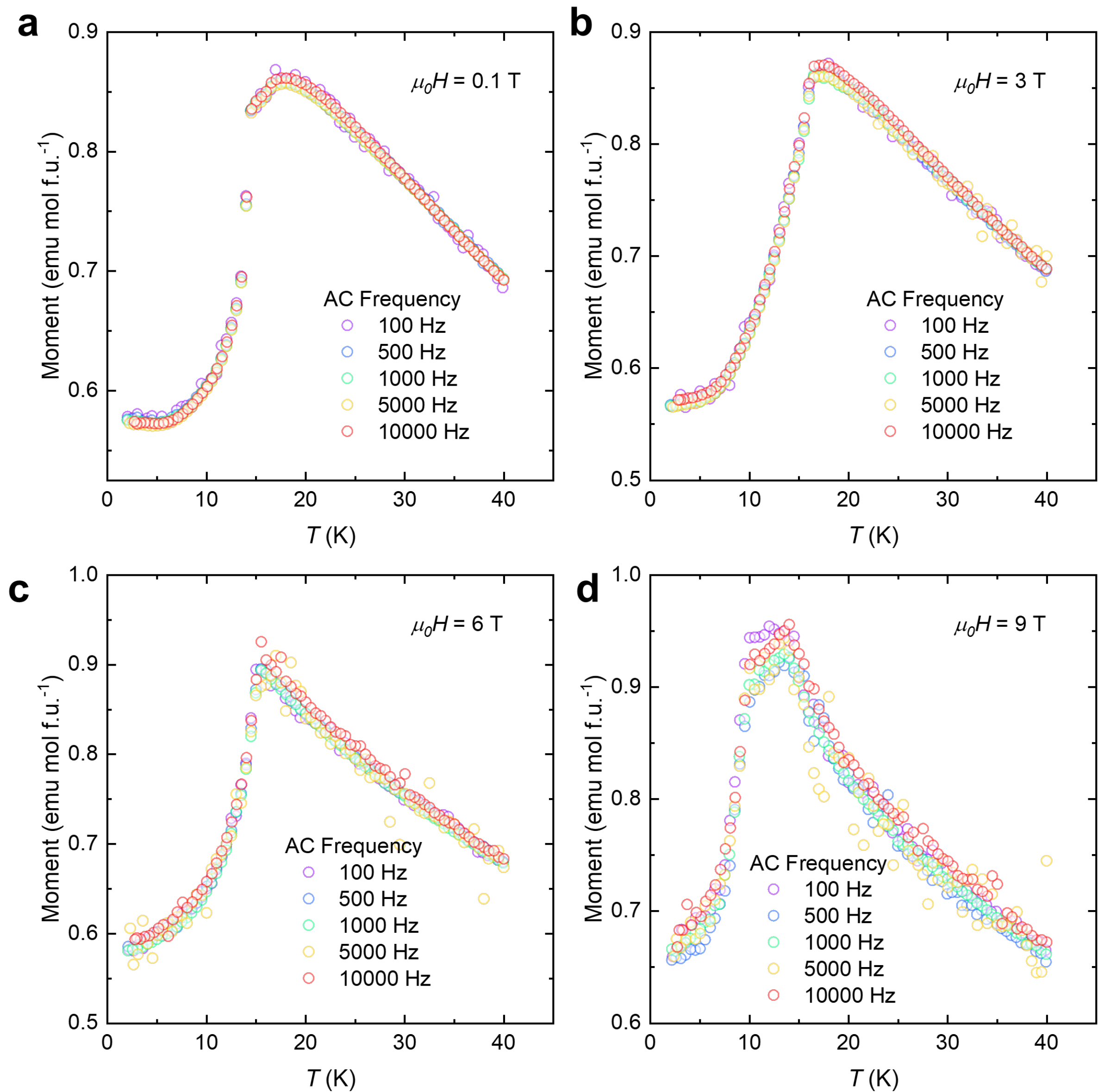

**Figure S7.** AC magnetization at around transition temperature under different frequency.

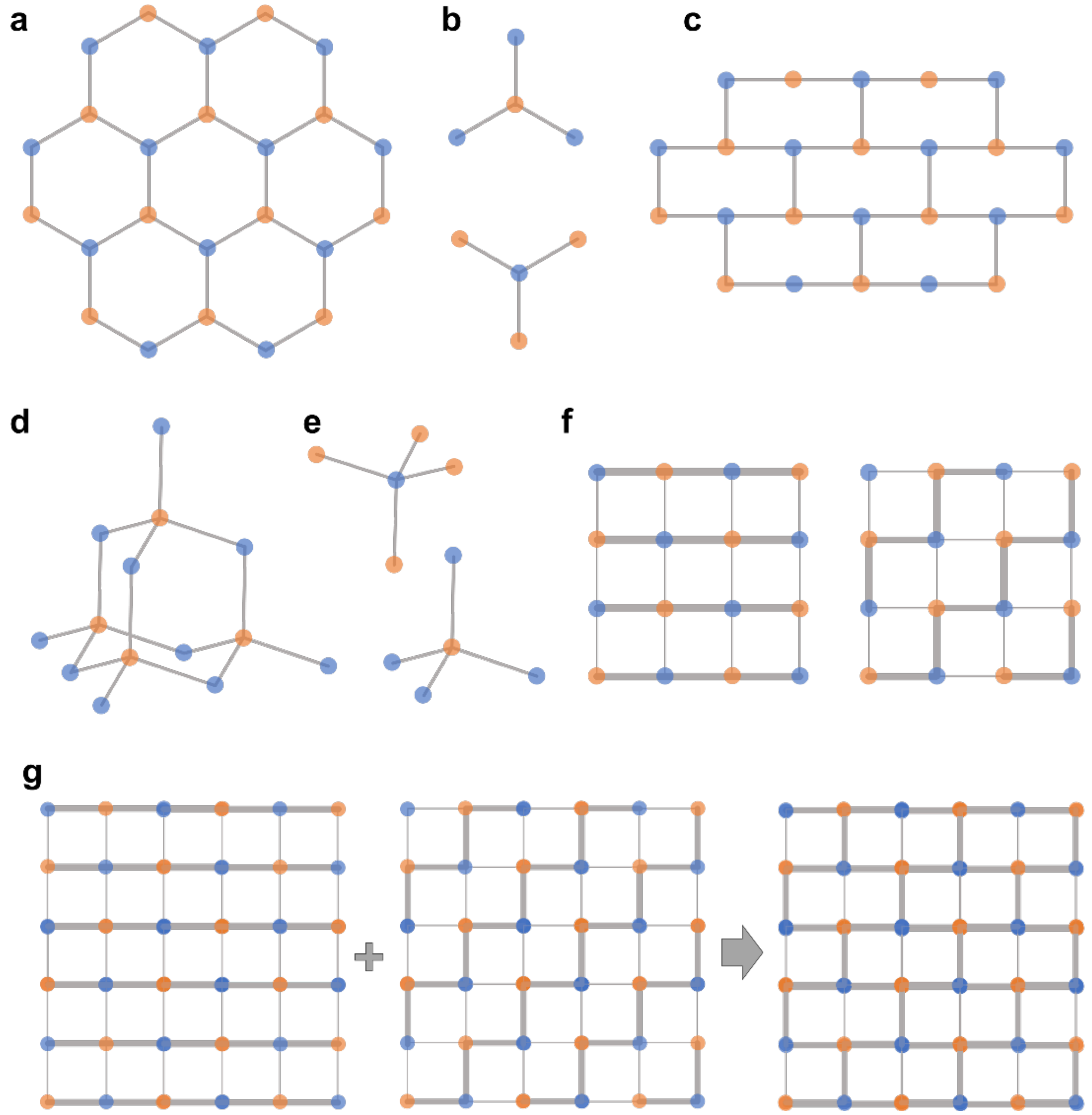

**Figure S8.** Comparison between honeycomb and diamond lattice.

The honeycomb lattice (a) and diamond lattice (d) are both bipartite lattices where the lattice is composed of two different sites, as shown in (b) and (e), and both features competing nearest-neighbor (between different sites) and second nearest-neighbor interactions (between same sites). When we pull the honeycomb lattice horizontally, the interactions in the honeycomb lattice can be illustrated as rectangular blocks with the strongest nearest-neighbor interactions highlighted in gray (c). When the diamond lattice is pressed into 2D, it forms a squire net, as shown in (f). In the $CaCo_2TeO_6$ case, among the four nearest neighbors, two of the Co-Co interactions are stronger (highlighted as bold gray lines) than the other due to different exchange interaction pathways. Due to the reduction of dimension, the 3-D diamond lattice shows both types of interactions as illustrated in (g), resulting in a honeycomb-like (approximate Kitaev) interaction.

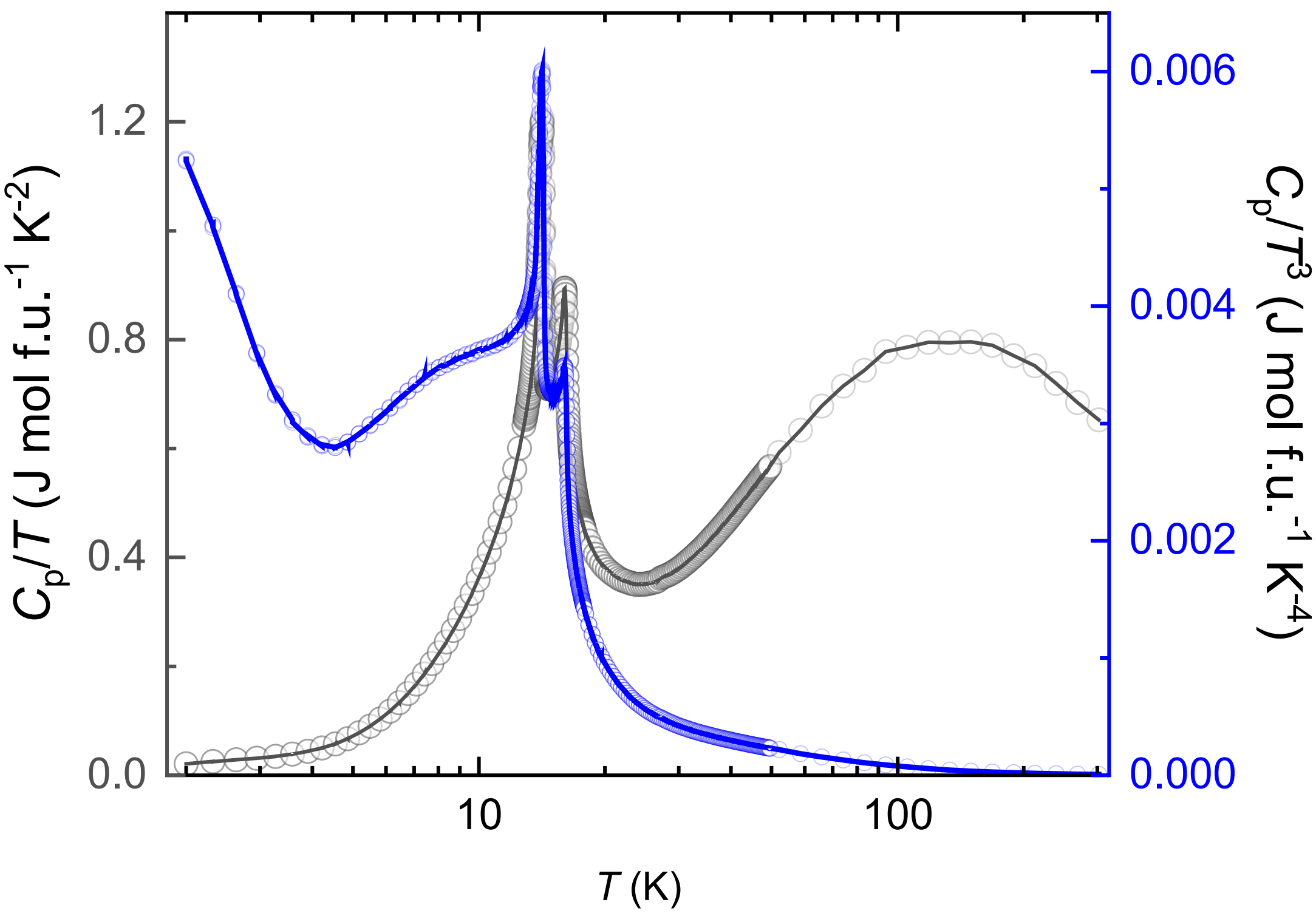

**Figure S9.** $C_p/T$ and $C_p/T^3$ as a function of temperature, showing the presence of Einstein mode phonon.

Extracting the magnetic contribution to the heat capacity requires a reliable phonon subtraction using a most direct nonmagnetic isostructural analog or a phonon model. We attempted to prepare the unknown $CaZn_2TeO_6$ as the nonmagnetic version for subtracting the lattice contribution, but were not successful. Thus, we constructed a phonon model that best describes the high-temperature heat capacity data by using two Debye modes and one Einstein model.

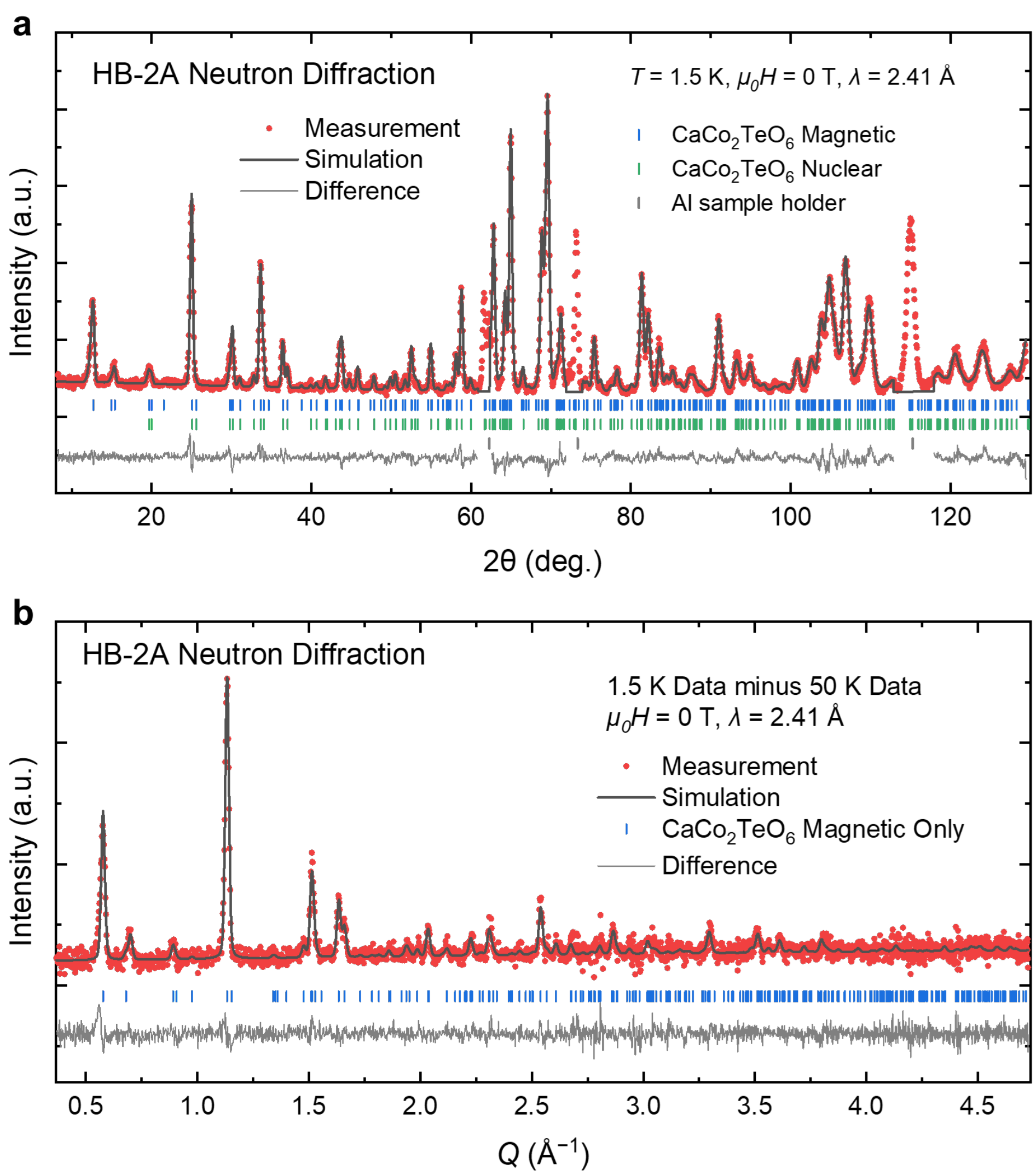

**Figure S10.** HB-2A Powder neutron diffraction.

The magnetic structure was fitted with all maximal magnetic space groups (As shown in the table below). The *Pnma* (62.441) shows the best agreement in terms of peak position and intensity.

| MSG | # Parameters | $R_{WP}$ | $\chi^2$ |
|---|---|---|---|
| Pnma (62.441) | 6 | 15.5 | 2.871 |
| Pn’ma (62.443) | 6 | 26.8 | 7.789 |
| Pnm’a (62.444) | 6 | 26.1 | 7.356 |
| Pnma’ (62.445) | 6 | 19.1 | 3.968 |
| Pn’m’a (62.446) | 6 | 27.7 | 8.315 |
| Pnm’a’ (62.447) | 6 | 22.6 | 5.515 |
| Pn’ma’ (62.448) | 6 | 23.7 | 6.069 |
| Pn’m’a’ (62.449) | 6 | 25.5 | 7.021 |

The magnetic vectors were initially allowed to refine freely, and the result shows a small value and large error for $M_x$ , indicating the $M_x$ parameters are not well constrained.

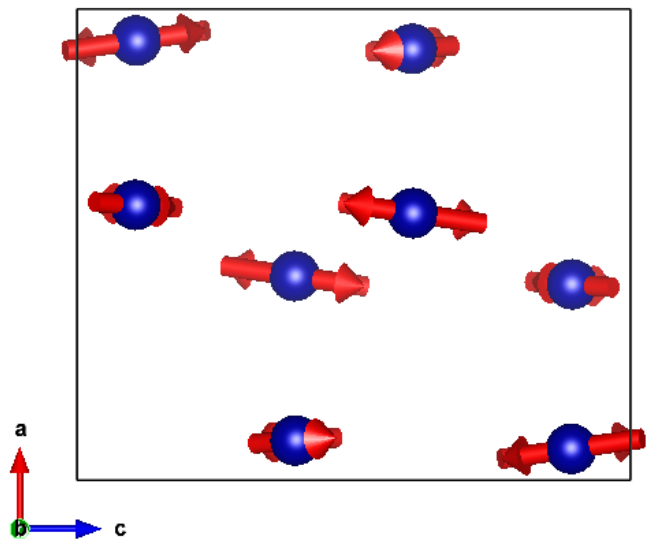

| Co | $M_x$ | $M_y$ | $M_z$ | $M_{tot}$ |
|---|---|---|---|---|
| Co(1) | -0.412(96) | -1.040(72) | 2.863(26) | 3.07 |
| Co(2) | -0.130(100) | 1.373(71) | 1.644(27) | 2.16 |

Then we tried to set the magnetic moment to zero.

| Co | $M_x$ | $M_y$ | $M_z$ | $M_{tot}$ |
|---|---|---|---|---|
| Co(1) | 0 (Fixed) | -1.045(74) | 2.862(26) | 3.04 |
| Co(2) | -0.477(84) | 1.371(73) | 1.663(27) | 2.21 |

| Co | $M_x$ | $M_y$ | $M_z$ | $M_{tot}$ |
|---|---|---|---|---|
| Co(1) | -0.458(77) | -1.072(85) | 2.863(26) | 3.09 |
| Co(2) | 0 (Fixed) | 1.345(84) | 1.659(27) | 2.14 |

And if both are fixed to 0:

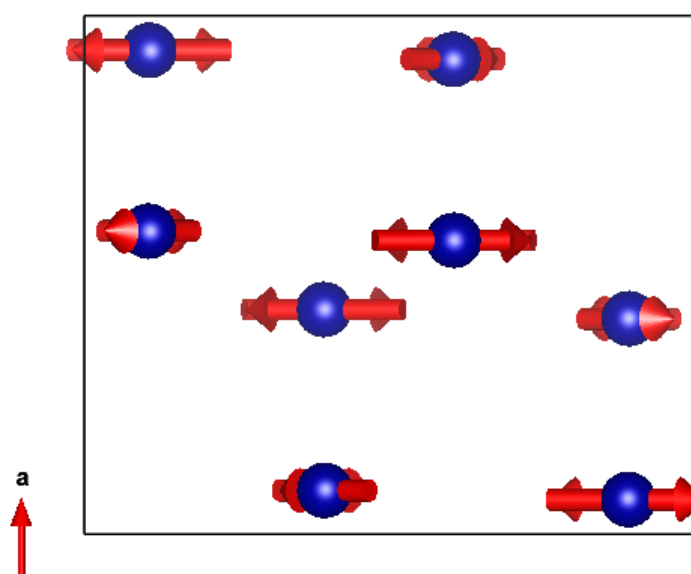

| Co | $M_x$ | $M_y$ | $M_z$ | $M_{tot}$ |
|---|---|---|---|---|
| Co(1) | 0 (Fixed) | -0.91(1) | 2.07(2) | 2.3(1) |
| Co(2) | 0 (Fixed) | 0.91(1) | 1.16(1) | 1.4(1) |

The diffraction data still fit well (Figure S22). Overall, slight canting out of the *bc*-plane can be included without affecting the fit drastically.

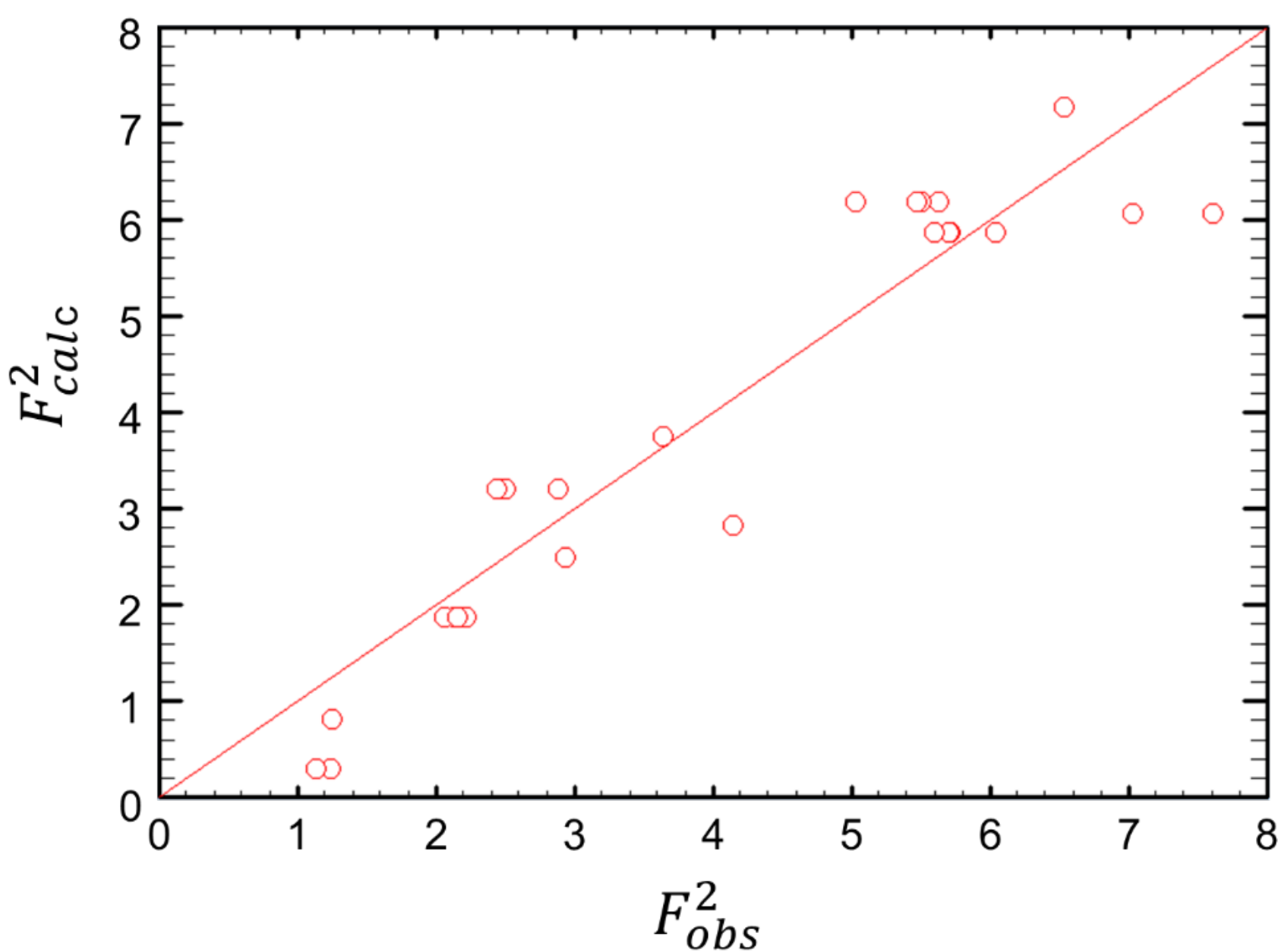

**Figure S11.** The observed magnetic squared structure factors vs. calculated from the FullProf refinement from CORELLI data.

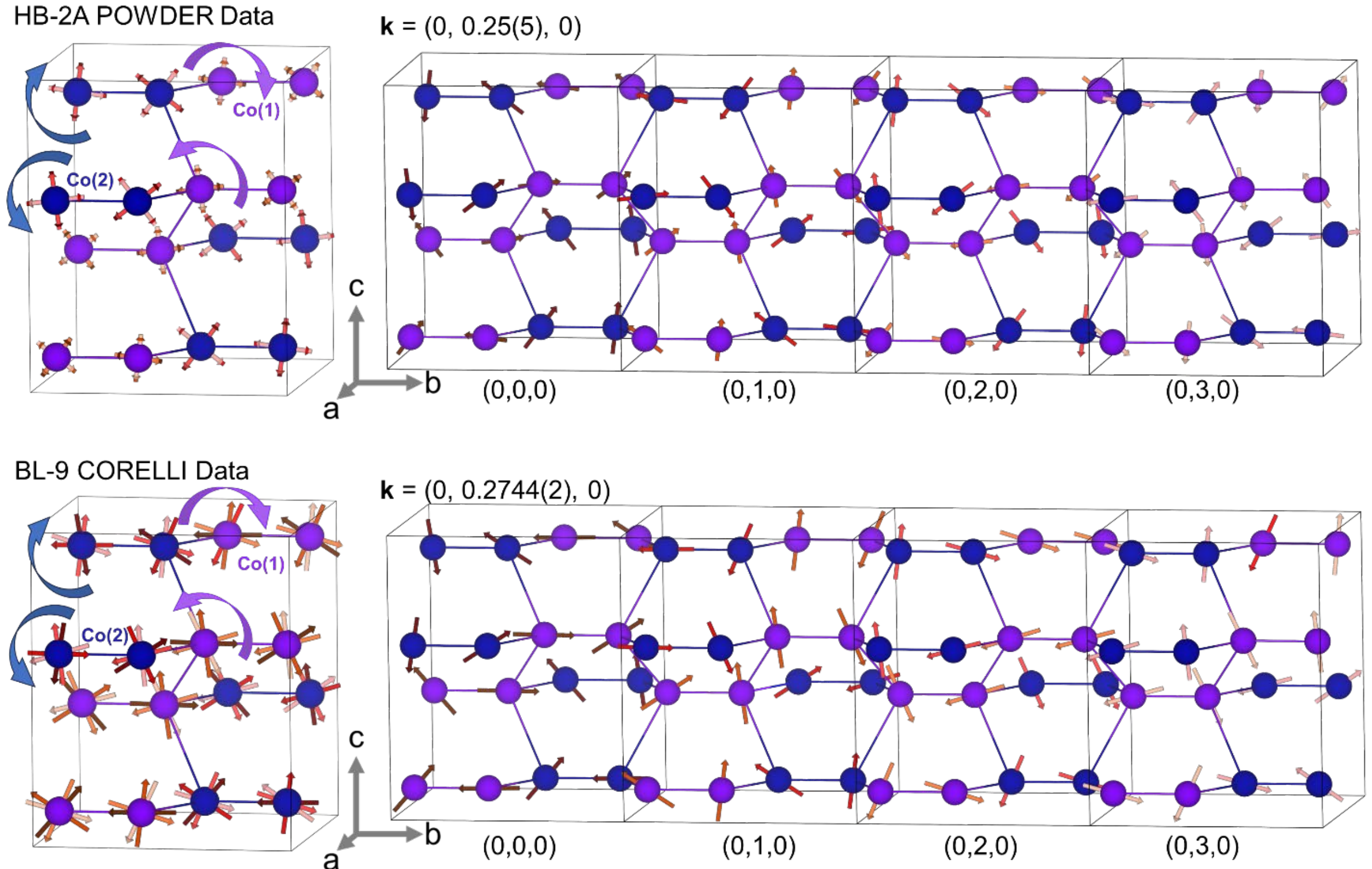

**Figure S12.** Magnetic structure at 15 K from HB-2A powder neutron diffraction and BL-9 single crystal neutron diffraction at 14.5 K.

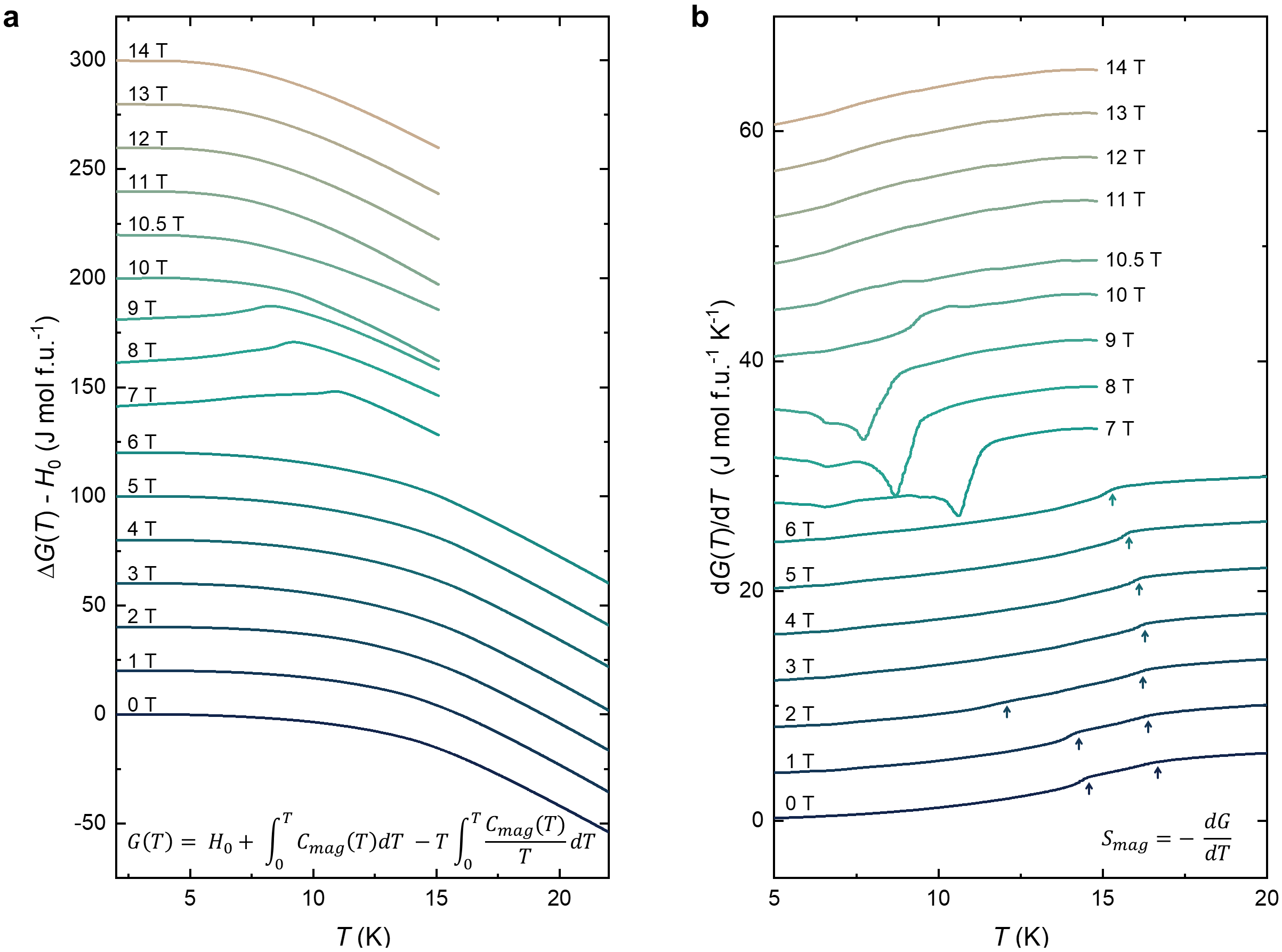

**Figure S13.** Magnetic Gibbs free energy and entropy as a function of temperature and magnetic field. (a) $\Delta G(T)-H_0$ obtained from $C_{\mathrm{mag}}(T)$ data. (b) Magnetic entropy $S_{\mathrm{mag}} = -\mathrm{d}G/\mathrm{d}T$.

To improve visibility, a constant vertical offset of +20 J mol$^{-1}$ was added sequentially to the Gibbs free energy curves for each higher-field dataset relative to the previous one, with the 0 T curve remaining unshifted. Similarly, a vertical offset of +4 J mol$^{-1}$ K$^{-1}$ was applied sequentially to the d$G(T)$/d$T$ curves between adjacent fields to enhance clarity.

To better understand *the nature* of the transitions presented in Fig 5h, we evaluated the magnetic Gibbs free energy and its temperature derivative, which corresponds to magnetic entropy (Fig. S13). Below 6 T, the entropy curves exhibit slight discontinuities and downward-bending features, and characteristic jump in entropy of first-order transitions associated with the PM → I-AFM → NC-AFM sequence. From 7 T to 10.5 T, the transitions remain first order, though the system no longer passes through the I-AFM state. In this field range, the unusual curvature in the entropy reflects proximity to a triple point where PM, spin-flop AFM (SF-AFM), and I-AFM phases intersect. The latent heat associated with the first order transition features are also shown in the heat curve (Fig. S14) as shoulders in the curve at $\mu_0 H \leq 10.5$ T. While at above 11 T, the entropy evolves smoothly and bends upward (Fig. S13), and the heat curve shows absence of latent heat, indicating a crossover to second-order behavior as the system enters a quantum paramagnetic (QPM) phase. These findings are consistent with the metamagnetic transitions and enhanced entropy observed in magnetization and heat capacity data (Figs. 2e,f and S7).

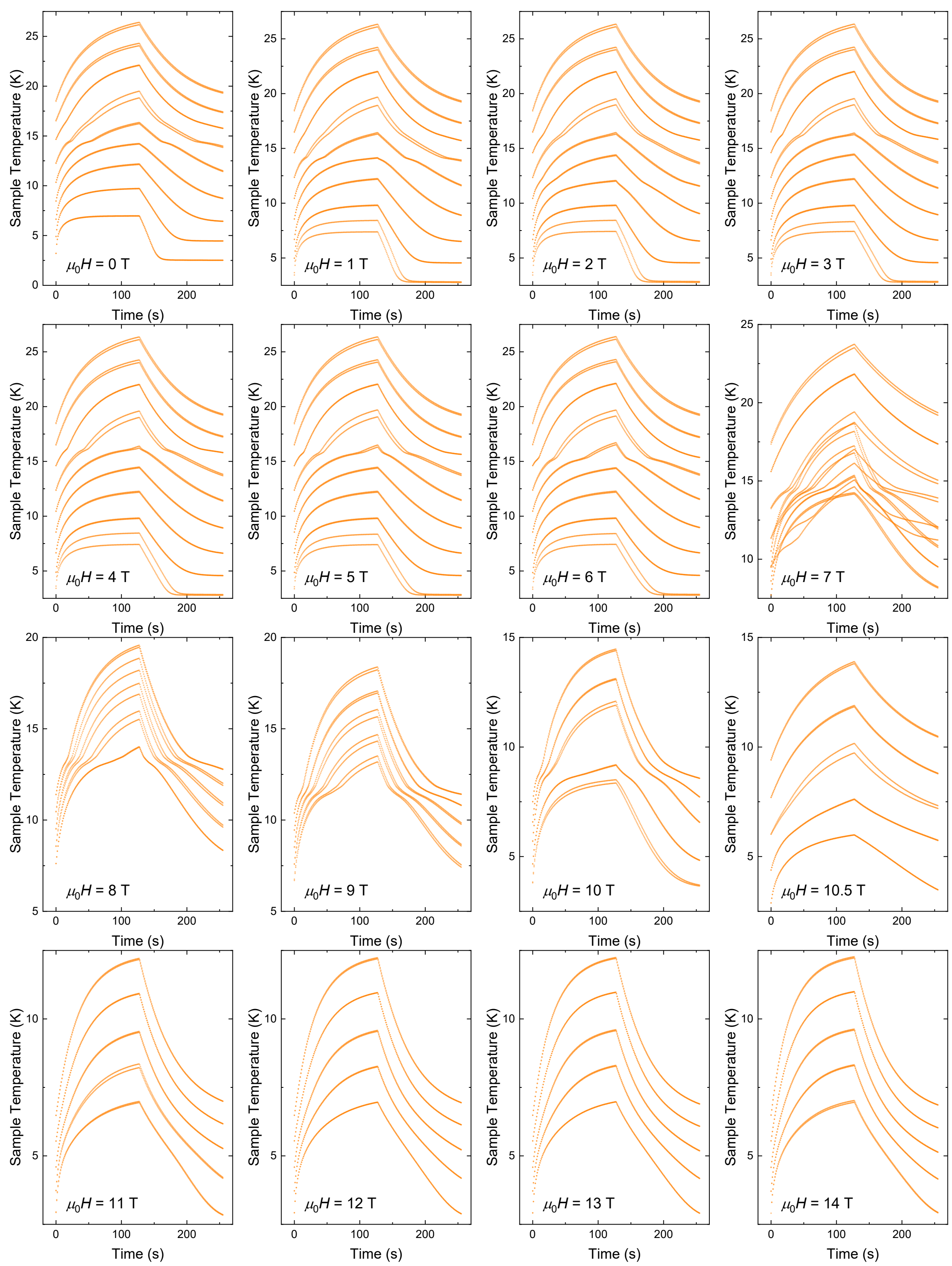

**Figure S14.** Heat curves from temperature dependent heat capacity measurements under fields.

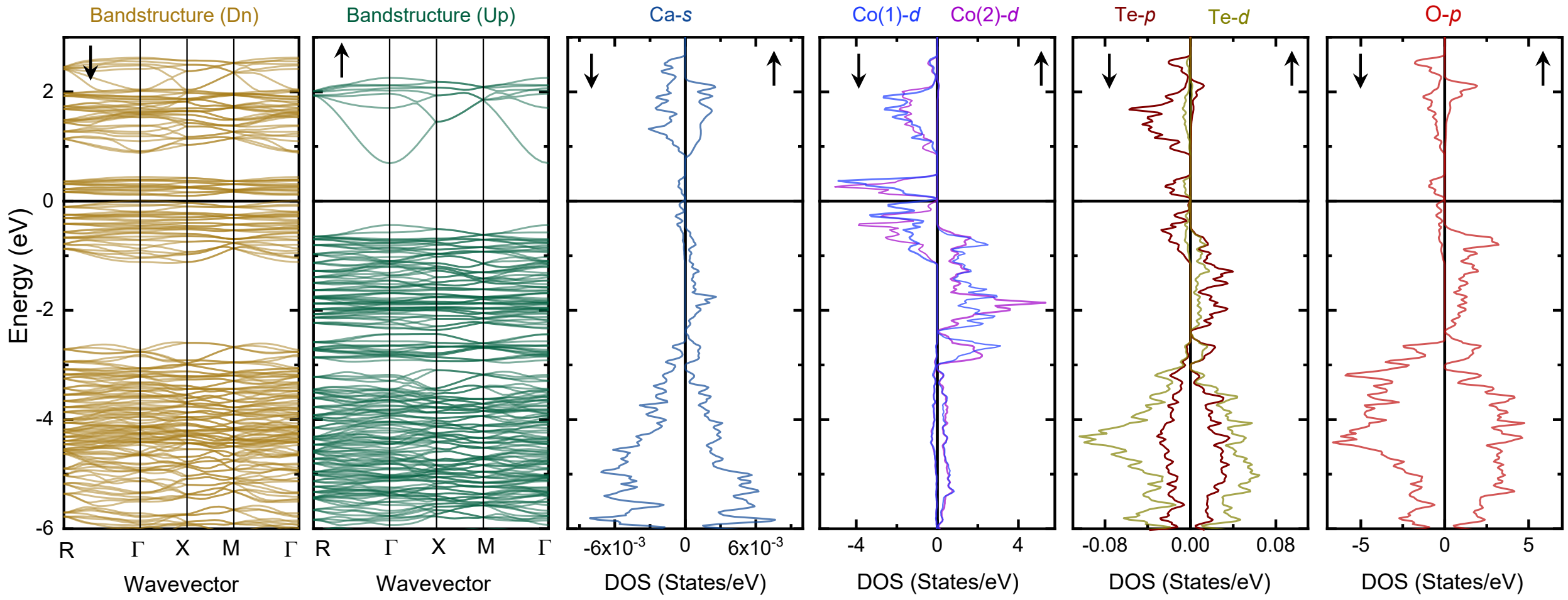

**Figure S15.** FP-LAPW band structure and DOS of $CaCo_2TeO_6$.

To connect the quantum-mechanical interference phenomena to the uniqueness of $CaCo_2TeO_6$ full-potential linearized augmented plane-wave (FP-LAPW) spin-polarized DFT calculations were performed (Fig. S15). The bands around the Fermi level ($E_F$) are diffused, indicating covalent characters and appreciable overlap between the Co-*d* and O-*p* states. The spins of the Co-*d* states are polarized and further polarize the O-*p*, Te-*s*/*p*, and Ca-*s* states, suggesting Heisenberg magnetic interactions along different directions. The valence band maximum and conduction band minimum primarily consist of the Co-*d* and O-*p* states. This indicates sizable Co-O-Co interactions, facilitating magnetic exchange pathways. The DFT results underscore the rich magnetic behavior of $CaCo_2TeO_6$ while appreciating the presence of competing nearest-neighbor and next-nearest-neighbor magnetic exchange interactions.

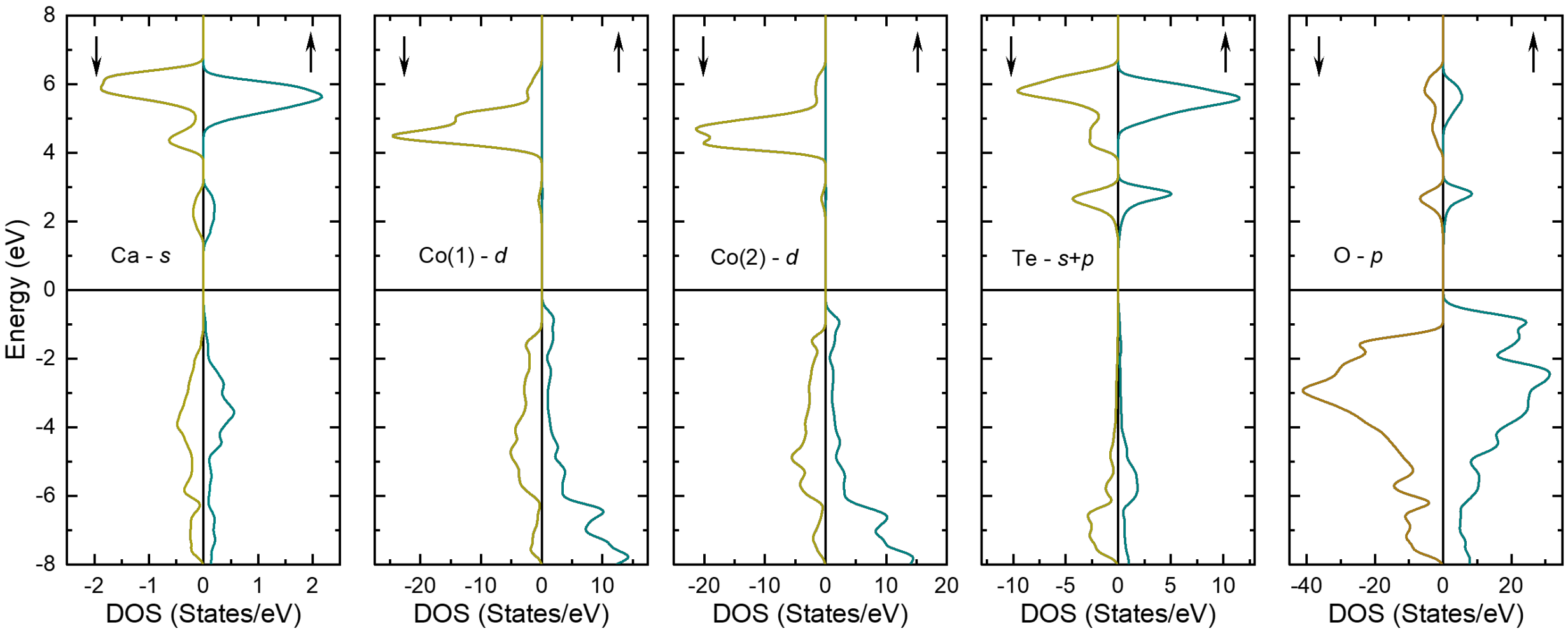

**Figure S16.** DOS of $CaCo_2TeO_6$ calculated using the PAW method.

Additional plane-wave DFT calculations were performed by utilizing the projector-augmented wave (PAW) method and projected into a linear combination of atomic orbital (LCAO). The spin-polarized DOS curves from the pseudopotential calculations are similar to those from FP-LAPW validating the DFT results.

To compare the overlap of the atomic interacting wavefunctions and the electronic instability of $T_d$-[$CoO_4$] in $CoRh_2O_4$ vs. $O_h$-[$CoO_6$] in $CaCo_2TeO_6$, we extracted the projected crystal orbital Hamilton population (-pCOHP) and integrated the COHP (ICOHP) of the Co-*d* and O-*p* orbitals within each $O_h$-[$CoO_6$] and $T_d$-[$CoO_4$] cluster up to $E_F$ per unit cell (Fig S16-21).

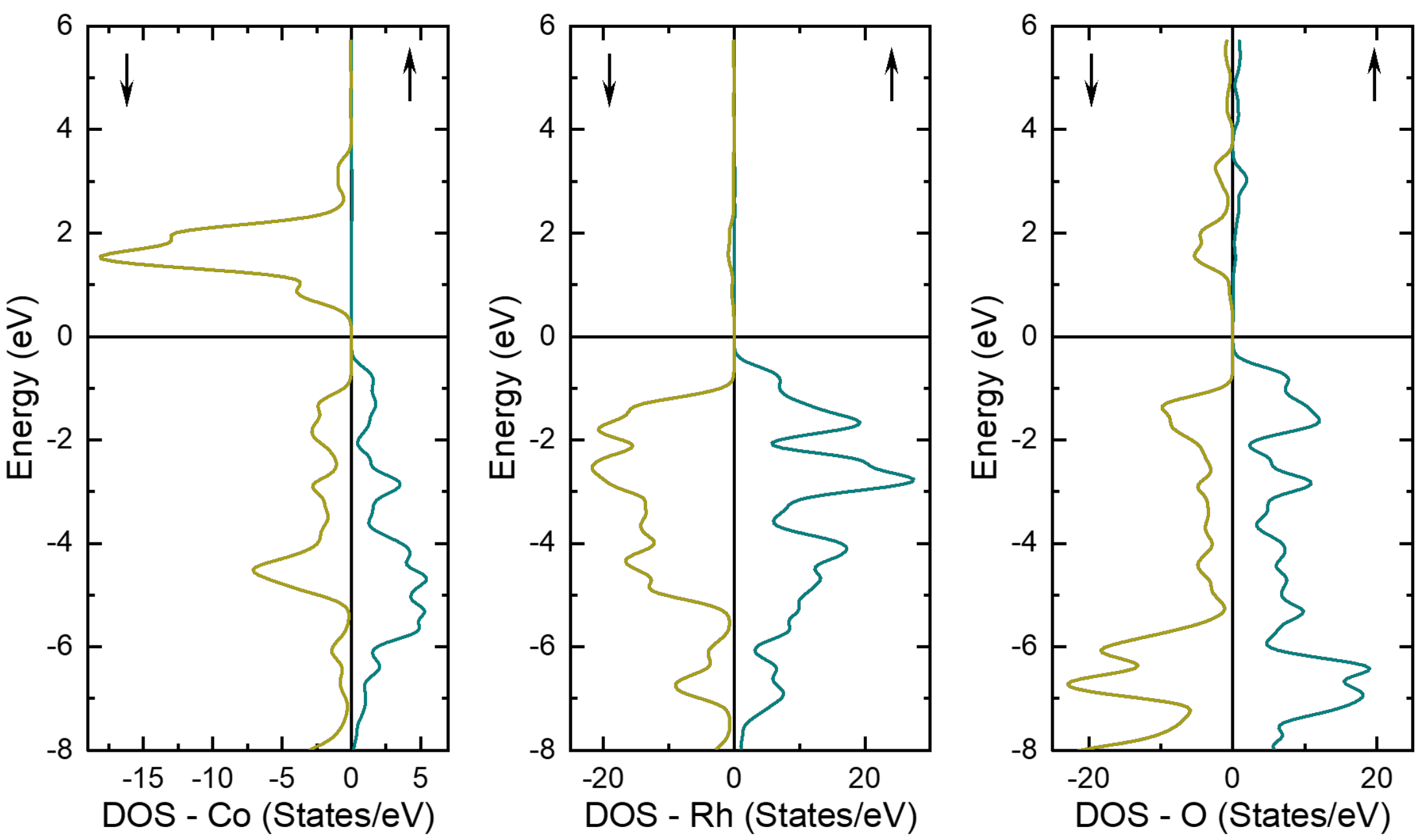

**Figure S17.** DOS of $CoRh_2O_4$ calculated using the PAW method.

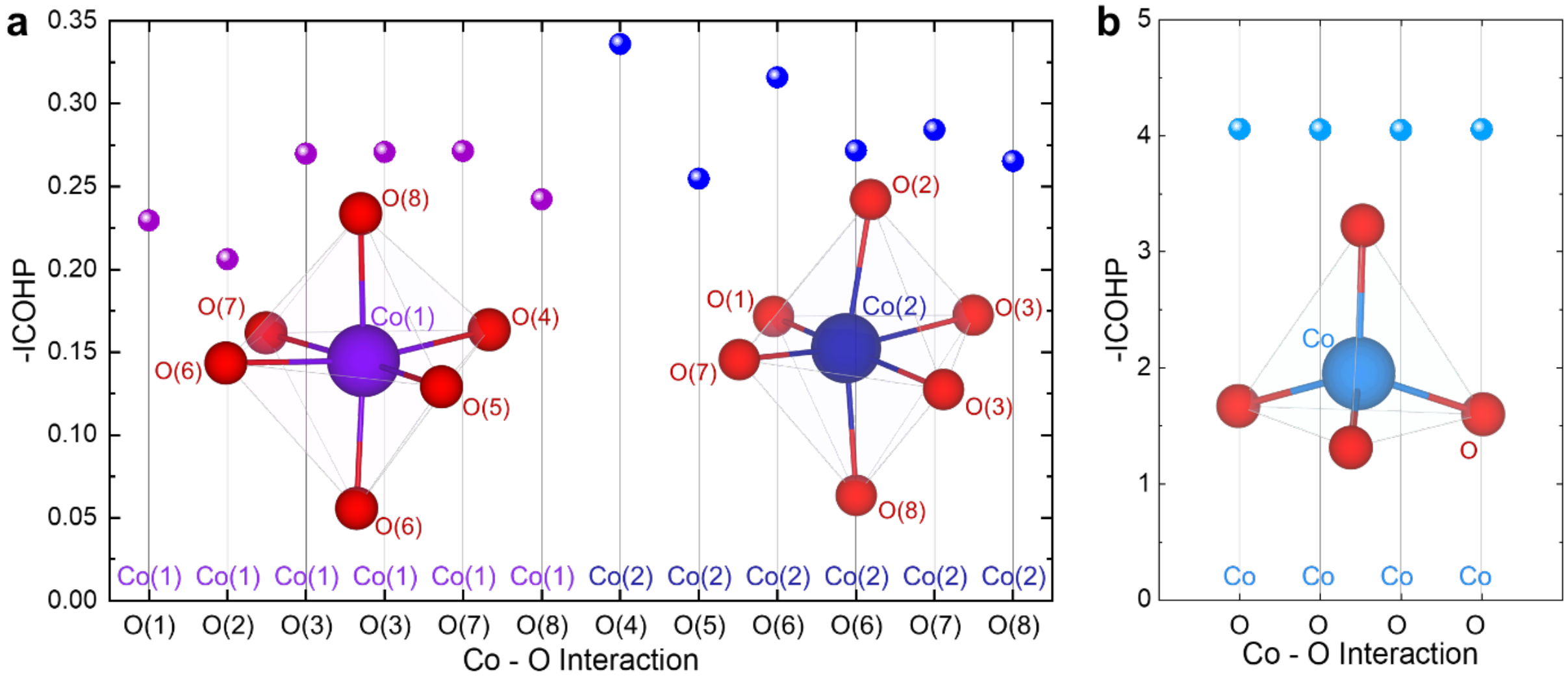

**Figure S18.** Crystal orbital Hamilton population and integration for Co-O bonds in (a) $CaCo_2TeO_6$ and (b) $CoRh_2O_6$.

The total integrated Co-O COHPs up to $E_F$ (ICOHPs) show overall bonding strength for $O_h$-[$CoO_6$] in $CaCo_2TeO_6$ and $T_d$-[$CoO_4$] in $CoRh_2O_4$. For $O_h$-[$CoO_6$], the total Co-O interaction strengths are similar in magnitude, and overall, the Co(1)-O bonds are slightly stronger than the Co(2)-O bonds. For $T_d$-[$CoO_4$], the Co-O bonds feature the same strength due to symmetry.

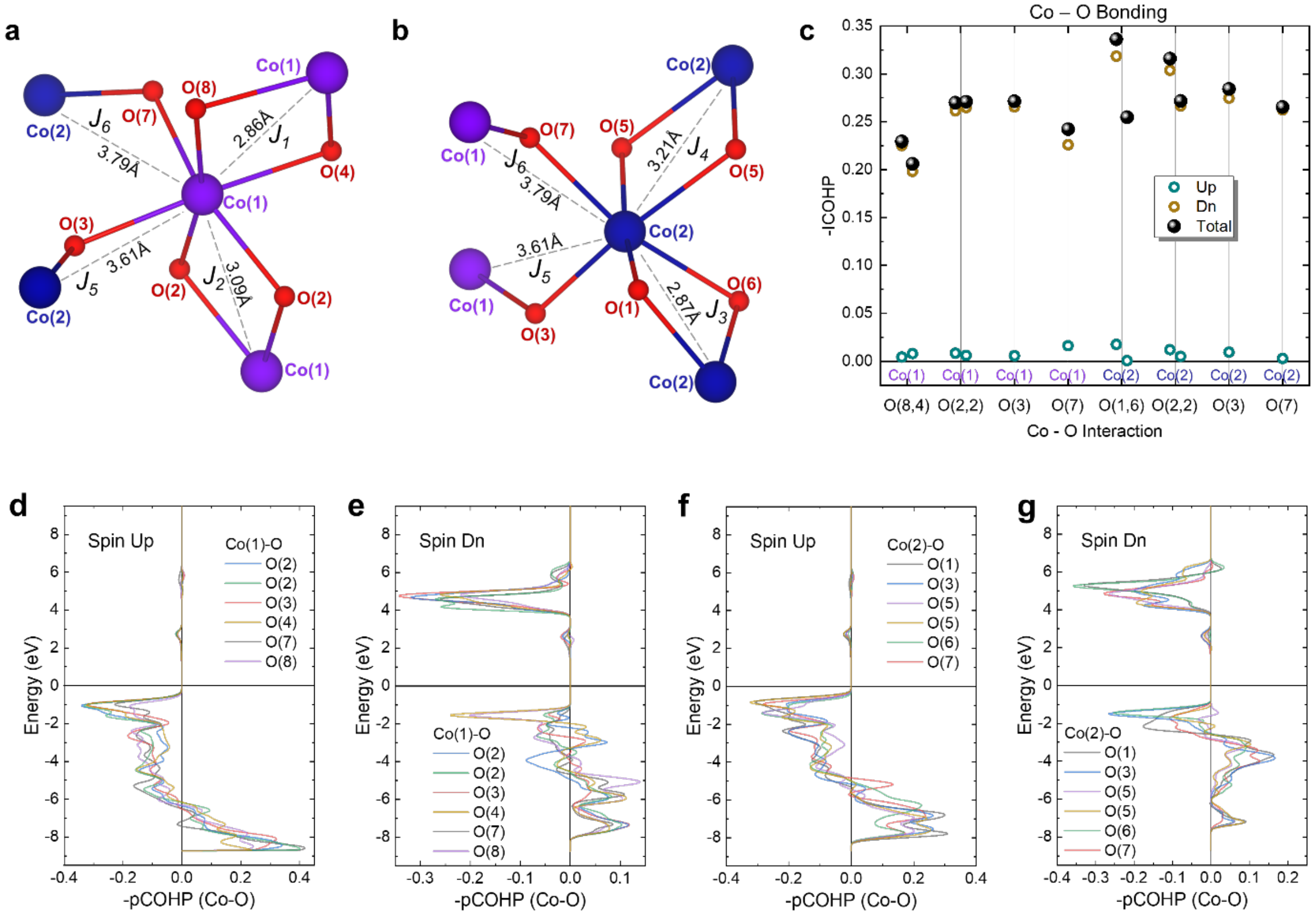

**Figure S19.** Crystal orbital Hamilton population and integration for Co-O bonds in $CaCo_2TeO_6$. The -pCOHP curves for $O_h$-[$CoO_6$] shows Co-O antibonding characters, as proposed in Fig. S1.

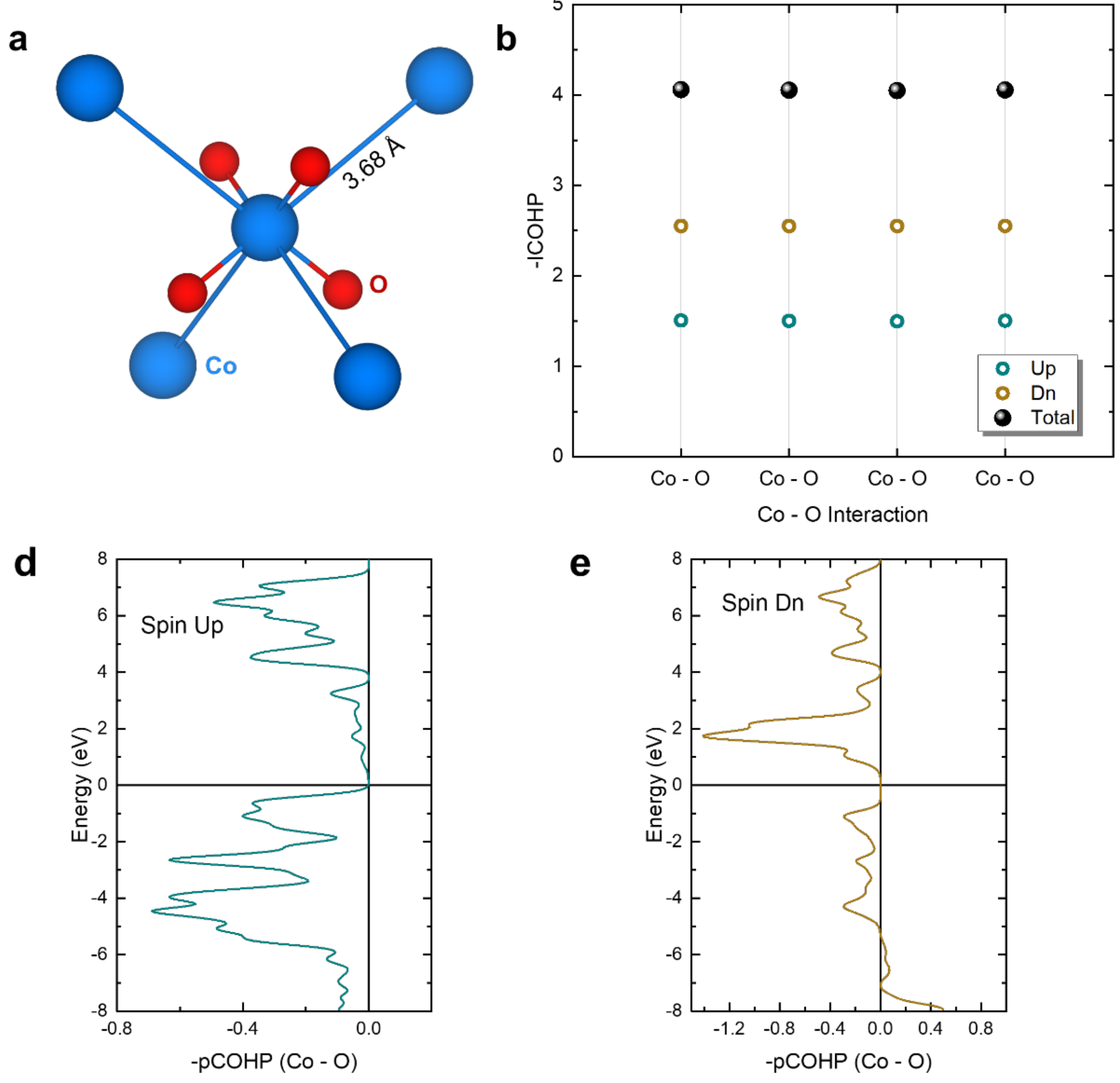

**Figure S20.** Crystal orbital Hamilton population and integration for Co-O bonds in $CoRh_2O_6$. The -pCOHP curves for $T_d$-[$CoO_4$] also shows Co-O antibonding characters.

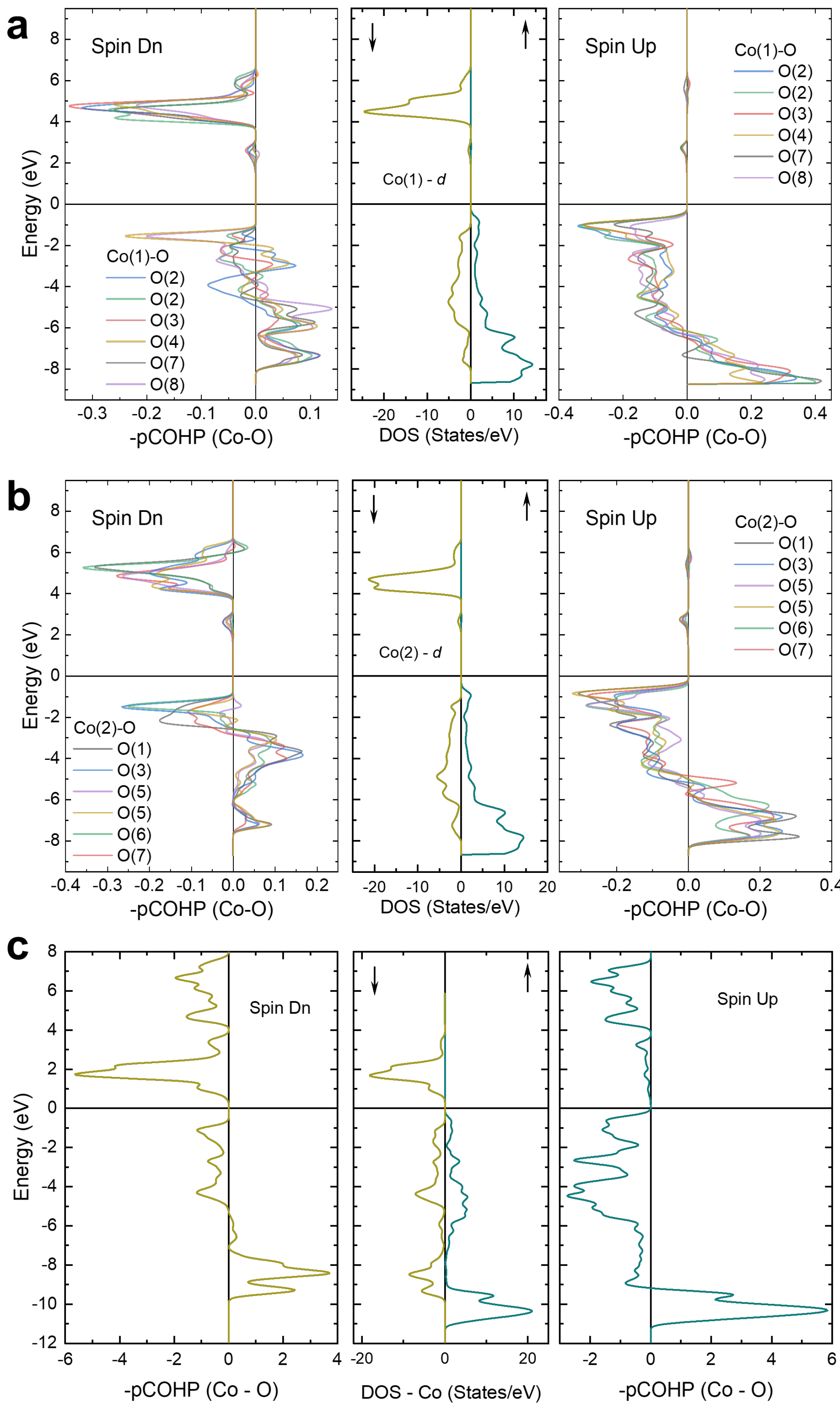

**Figure S21.** DOS of Co-*d* and pCOHP.
The energy scale integrated in Fig. 5a, b is determined from the DOS curve.

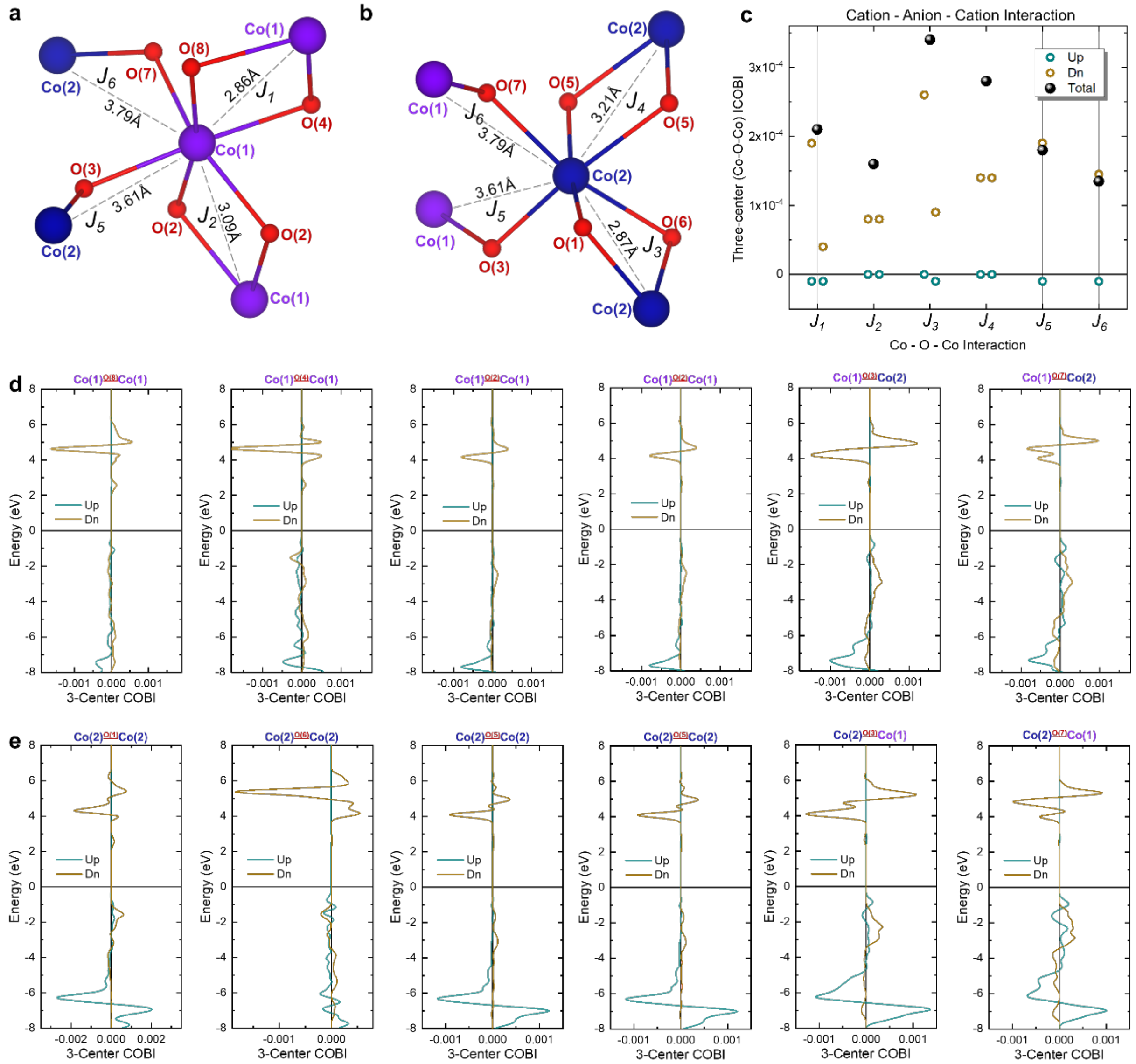

**Figure S22.** Crystal Orbital Bond Index for Co-O-Co bonds in $CaCo_2TeO_6$.

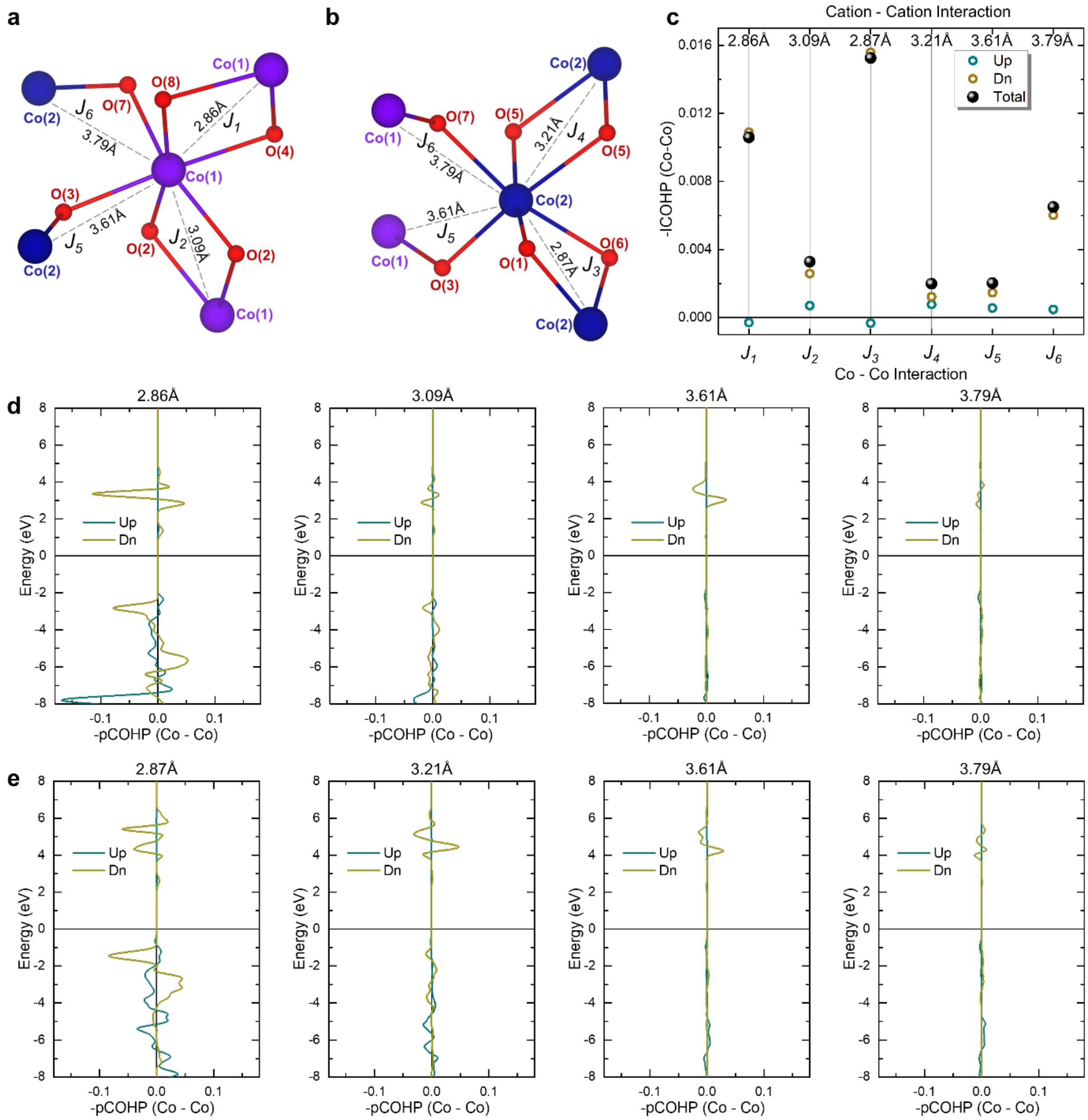

**Figure S23.** Crystal orbital Hamilton population and integration for Co-Co interactions in $CaCo_2TeO_6$.

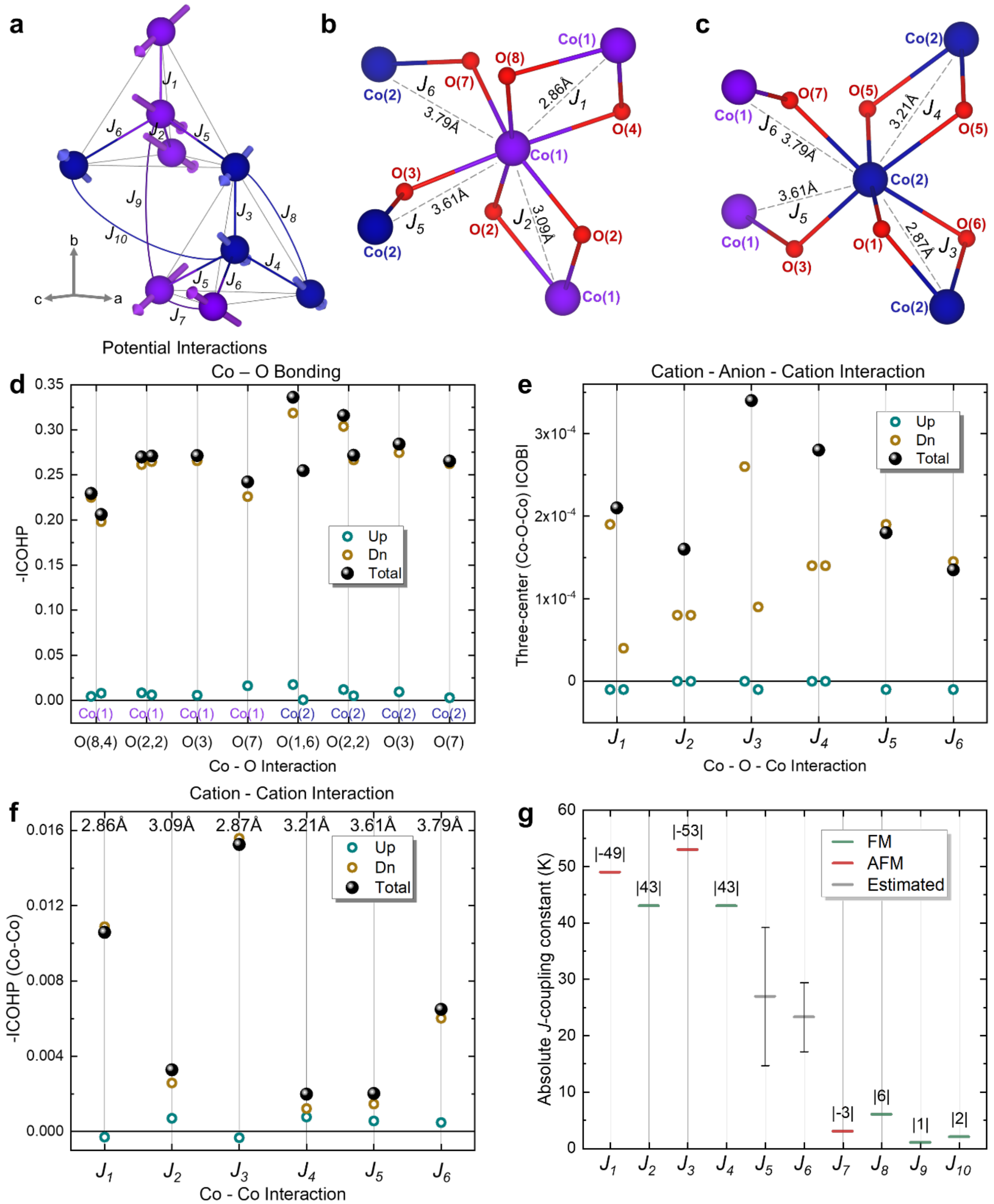

**Figure S24.** Interactions in $CaCo_2TeO_6$.

The three-center COBI curves (Fig. S22) between same pairs of Co atoms (e.g. Co(1)$\overset{\mathrm{O(8,4)}}{\text{——}}$Co(1) and Co(2)$\overset{\mathrm{O(1,6)}}{\text{——}}$Co(2) curves) show similar features that confirmed that the bond strength between Co and O are comparable. And the integration (ICOBI) (Fig. S22c) implies that interactions between identical Co types ($J_{1-4}$, facilitated through two Co-O-Co pathways: Co(1)$\overset{\mathrm{O(8,4;\ 2,2)}}{\text{———}}$Co(1), Co(2)$\overset{\mathrm{O(1,6;\ 5,5)}}{\text{———}}$Co(2)) surpass those between different Co types ($J_{5,6}$, mediated by a singular O bridge: Co(1)$\overset{O(3)}{\text{——}}$Co(2), Co(1)$\overset{O(7)}{\text{——}}$Co(2)). Consistent with the ICOHP, the cation-anion-cation interaction between Co(2) is still stronger than that for Co(1). Although the total ICOBI between Co(1)$\overset{O(2,2)}{\text{——}}$Co(1) and Co(1)$\overset{O(3)}{\text{——}}$Co(2) appears identical, the 3-center ICOBI only take the cation - anion – cation ($e_g$ – p - $e_g$) into account, ignoring the direct cation-cation ($t_{2g}$ - $t_{2g}$) interaction between edge-shared Co atoms.[12, 13] To quantify the direct interactions between the same type of Co, we calculated the COHP between Co in the shown cluster (Fig. S23). The ICOHP (Fig. S23f) confirmed that the interaction between Co(1)-Co(1) (3.09Å) is stronger than that of Co(1)-Co(2) (3.61Å). Moreover, if we take a closer look at the cation-cation interaction, the Co(1)$\overset{O(8,4)}{\text{——}}$Co(1) and Co(2)$\overset{O(1,6)}{\text{——}}$Co(2) interactions are a lot stronger than that in Co(1)$\overset{O(2,2)}{\text{——}}$Co(1) and Co(2)$\overset{O(5,5)}{\text{——}}$Co(2) although they both form edge sharing octahedra. This is most likely due to the local environment difference (Fig. S25a). Take Co(1) as an example, the bonds between Co(1)-O(4)-Co(1)-O(8) are perpendicular to the $d$-$z^2$ orbital, while the Co(1)-O(2)-Co(1)-O(2) bonds are distorted due to crystal field splitting and deviate from the $d$-$z^2$ orbital, so the $d$-$e_g$ orbitals of Co(1)$\overset{O(2,2)}{\text{——}}$Co(1) are pointing at similar direction but different layer, resulting in less overlap. This direct cation-cation interactions can be confirmed from the DOS curve of decomposed $d$-$t_{2g}$ orbitals (Fig. S25b,c), that the $d$-yz and $d$-xz orbital in spin up, $d$-xy and $d$-xz in spin down states from Co(1), and $d$-xy and $d$-xz in spin up, $d$-yz and $d$-xz orbital in spin down states from Co(2) falls to similar trends. In addition, the Co(1)-Co(2) (3.61Å) interaction is weaker than Co(1)-Co(2) (3.79Å) interaction even though the atomic distance is closer. This could be arisen from the potential $e_g$-$e_g$ interaction between different types of Co (Fig. S26a), where the Co(1)-O(7)-Co(2) bond angle (127.9(1)$^o$) is closer to 180$^o$ than Co(1)-O(3)-Co(2) (119.2(1)$^o$). Indicating that the $d$-$e_g$ orbitals of site d and e are pointing at similar directions. The decomposed $d$-$e_g$ orbitals (Fig. S26b, c) confirmed this hypothesis that the $d$-$z^2$ and $d$-$x^2$-$y^2$ orbitals from different Co sites follows the same trends.

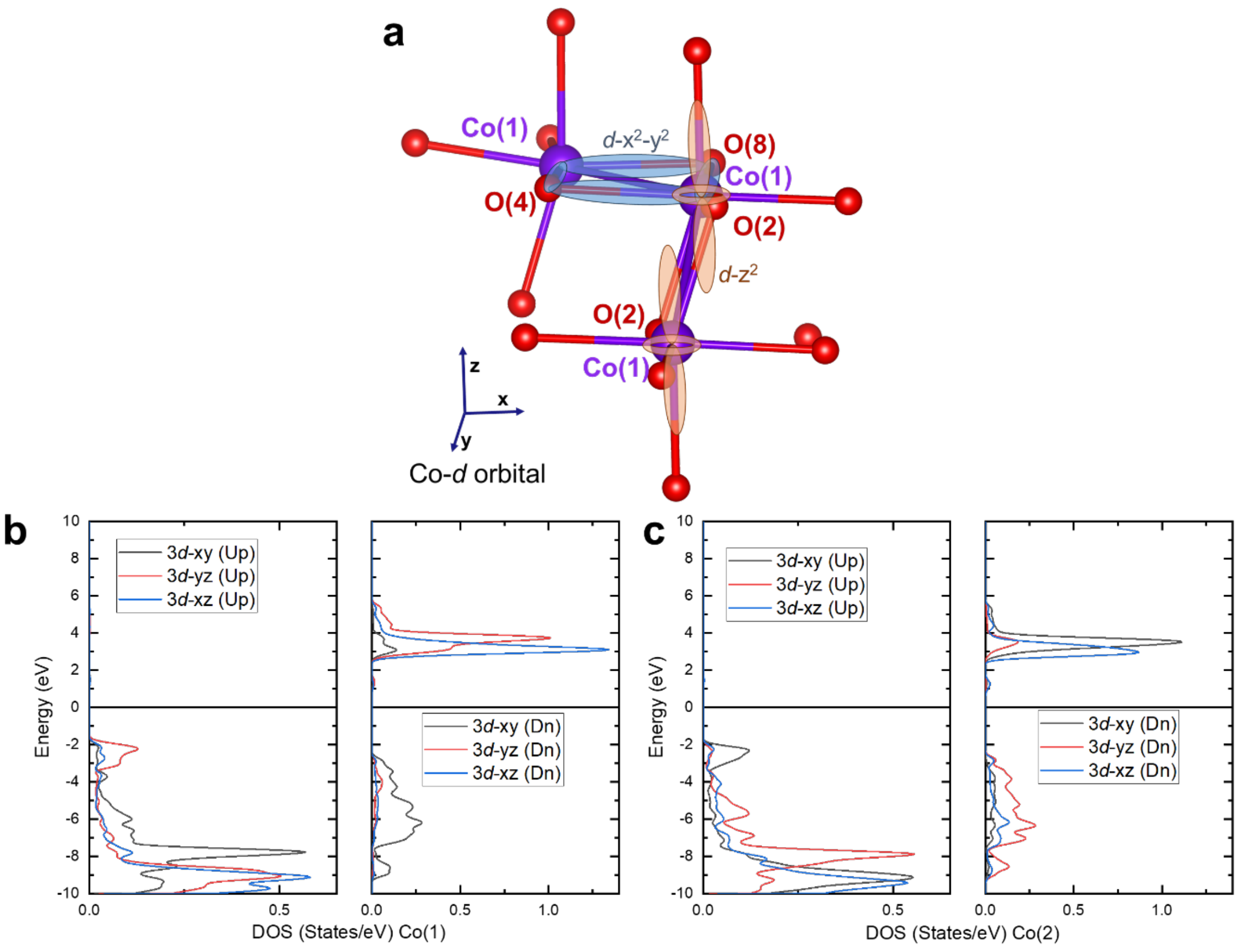

**Figure S25.** Local environment in Co(1)-centered tetrahedra between same type of Co.

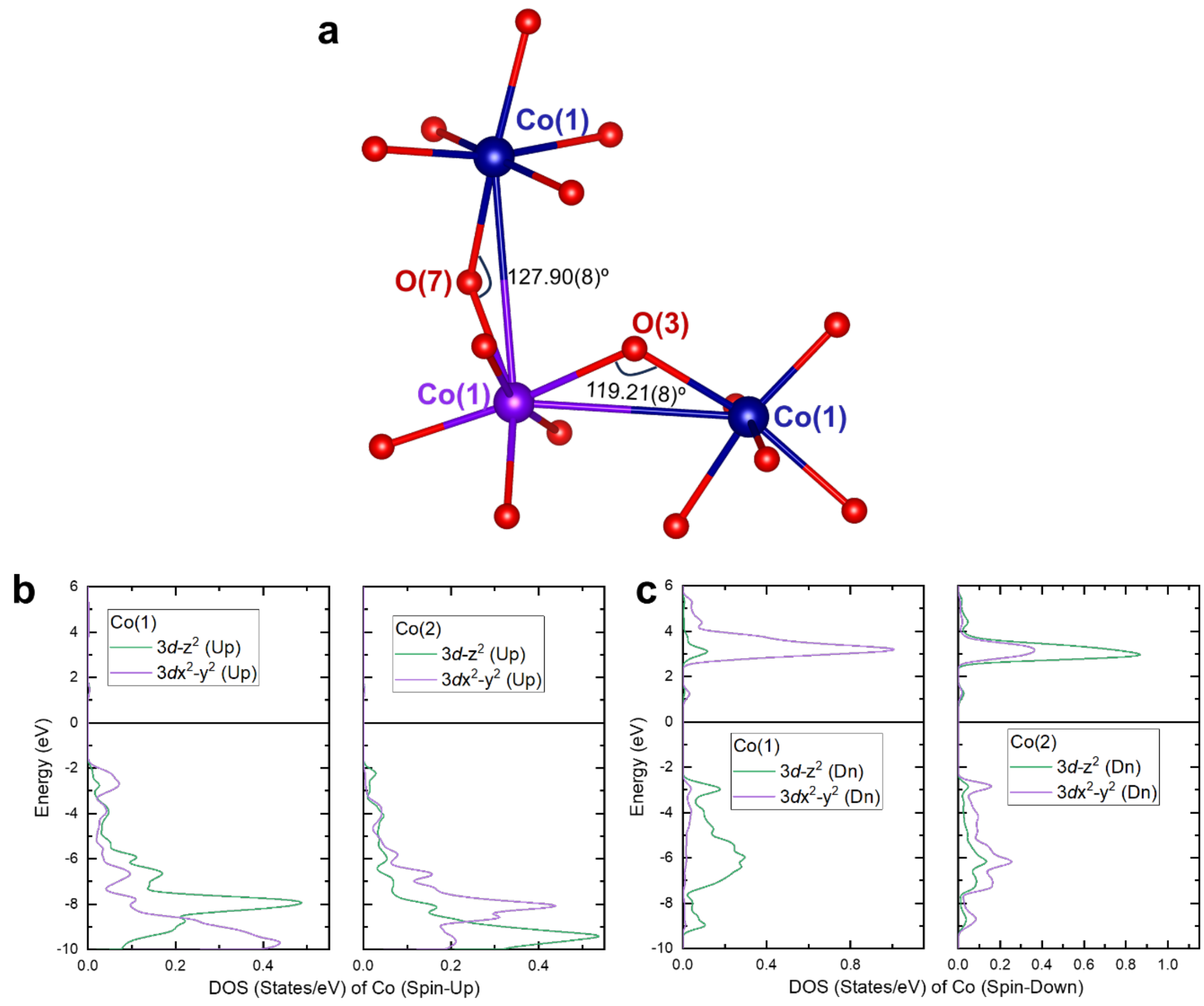

**Figure S26.** Local environment in Co(1)-centered tetrahedra between different types of Co.

**Table S1.** $CaCo_2TeO_6$ and compounds with similar magnetic structure.

| **Compound** | **Space Group** | $T_N$ **(K)** | **Magnetic ground state** | **Observed Moment (μB)** | **Expected Moment (μB)** | **Moment Recovery (%)** | **Reason for reduction** |
|---|---|---|---|---|---|---|---|
| $MnSc_2S_4$ [14] | *Fd*-3*m* | 2 | Spin-liquid above $T_N$, antiferromagnetic below $T_N$ | N/A | 5.92 | N/A | Frustration and spin-liquid behavior |
| $FeSc_2S_4$ [14] | *Fd*-3*m* | <0.05 | Spin-orbital liquid (no long-range order) | N/A | 4.9 | N/A | Spin-orbital liquid state |
| $CoRh_2O_4$ [15] | *Fd*-3*m* | 25 | Collinear Néel ordered state | 2.4(1) | 3.87 | 63.5 | Spin-wave corrections and quantum effects |
| $LiYbO_2$ [16] | $I4_1/amd$ | <1 | Incommensurate spiral, transitions to commensurate k=(1/3,±1/3,0) | 1.26 | 1.5 | 84 | Spiral ground state with frustration |
| $NaCeO_2$ [17] | $I4_1/amd$ | 3.18 | A-type antiferromagnetism with Jeff=1/2 moments | 0.57 | 2.14 | 26.6 | Crystalline electric field effects and Jeff=1/2 state |
| $CuRh_2O_4$ [15] | $I4_1/amd$ | 24 | Incommensurate helical order with reduced ordered moments | 0.47(5) | 1.0 | 47 | Quantum fluctuations and frustration |
| $KRuO_4$ [18] | $I4_1/a$ | 22.4 | Collinear antiferromagnet, weak frustration | 0.57(7) | ~1.0 | 57 | Spin-orbit coupling and short-range spin correlations |
| $TbTaO_4$ [19] | *I*2/*a* | 2.25 | Commensurate antiferromagnetic order | 7.5 (5) | 9.72 | 0 - 77.1 | Frustration and low-dimensional interactions |
| $CaCo_2TeO_6$ | *Pmna* | 17.1 | Collinear antiferromagnet, spin fluctuations | 3.03(1) and 2.15(1) | 3.36 | 90 and 64 | Competing interactions and quantum fluctuation |

- $J_1$ dominates over $J_2$ in most cases, indicating that nearest-neighbor coupling is the primary driver of magnetic behavior.
- Large $J_1/J_2$ ratios (e.g., >8 for $LiYbO_2$, 25 for $KRuO_4$) suggest well-defined ground states with relatively weak frustration (e.g., $CoRh_2O_4$) favor classical Néel ordering.
- Significant next-nearest-neighbor interactions (J2) lead to competing magnetic orders (e.g., spiral or helical ground states in $LiYbO_2$ and $CuRh_2O_4$).

**Table S2.** Exchange interactions.

| Centering Co | Distance to the same type of Co (Å) | | Distance to different types of Co (Å) | |
|---|---|---|---|---|
| Co(1) | 2.8627(6) ($J_1$) | 3.0922(7) ($J_2$) | 3.6131(5) ($J_5$) | 3.7901(5) ($J_6$) |
| Co(2 | 2.8672(6) ($J_3$) | 3.2139(7) ($J_4$) | 3.6131(5) ($J_5$) | 3.7901(5) ($J_6$) |

**Table S3.** Single-crystal, synchrotron XRD data refinement.

| Unit cell parameters | Single-crystal XRD | Synchrotron XRD |
|---|---|---|
| a (Å) | 9.2519(5) | 9.256351 (1) |
| b (Å) | 8.9784(5) | 8.991875 (1) |
| c (Å) | 10.8666(6) | 10.878668 (1) |

**Table S4.** Atomic coordinates.

| SXRD | x | y | z | U |
| --- | --- | --- | --- | --- |
| Ca | 0.75802(5) | 1.05122(5) | 0.34490(4) | 0.00426(10) |
| Co(1) | 0.56681(3) | 0.90942(3) | 0.60680(3) | 0.00306(8) |
| Co(2) | 0.58435(3) | 0.59033(3) | 0.10555(3) | 0.00277(8) |
| Te(1) | 0.43945(2) | 0.750000 | 0.37721(2) | 0.00203(7) |
| Te(2) | 0.58421(2) | 0.250000 | 0.10483(2) | 0.00192(7) |
| O(1) | 0.5062(2) | 0.750000 | 0.2123(2) | 0.0046(4) |
| O(2) | 0.57286(16) | 0.90733(16) | 0.41961(16) | 0.0037(3) |
| O(3) | 0.30110(17) | 0.59163(16) | 0.34416(15) | 0.0043(3) |
| O(4) | 0.4021(2) | 0.750000 | 0.5550(2) | 0.0040(4) |
| O(5) | 0.62146(17) | 0.40957(16) | -0.00876(15) | 0.0041(3) |
| O(6) | 0.3887(2) | 0.250000 | 0.0428(2) | 0.0036(4) |
| O(7) | 0.53159(17) | 0.40627(16) | 0.21923(15) | 0.0039(3) |
| O(8) | 0.7797(2) | 0.250000 | 0.1709(2) | 0.0042(4) |

| Synchrotron XRD | x | y | z | B |
| --- | --- | --- | --- | --- |
| Ca | 0.75802 | 1.05122 | 0.34490 | 1.22359 |
| Co(1) | 0.56685 | 0.90942 | 0.60924 | 0.97122 |
| Co(2) | 0.58483 | 0.59033 | 0.10588 | 0.66099 |
| Te(1) | 0.43700 | 0.75000 | 0.37867 | 0.65399 |
| Te(2) | 0.58208 | 0.25000 | 0.10590 | 0.78696 |
| O(1) | 0.50620 | 0.75000 | 0.21230 | 0.47481 |
| O(2) | 0.57286 | 0.90733 | 0.41961 | 0.33418 |
| O(3) | 0.30110 | 0.59163 | 0.34416 | 0.68840 |
| O(4) | 0.40210 | 0.75000 | 0.55500 | 0.40199 |
| O(5) | 0.62146 | 0.40957 | -0.00876 | 0.41520 |
| O(6) | 0.38870 | 0.25000 | 0.04280 | 0.67056 |
| O(7) | 0.53159 | 0.40627 | 0.21923 | 0.44008 |
| O(8) | 0.77970 | 0.25000 | 0.17090 | 0.41797 |

**Table S5.** Phonon estimation.

| Variable | Fitted value |
| --- | --- |
| $\gamma$ | 0.02671 ± 0.12987 |
| Number of oscillators of Debye mode 1 | 3.7244 ± 0.79847 |
| Debye temperature 1 | 393.85999 ± 39.62869 |
| Number of oscillators of Debye mode 2 | 5.46047 ± 1.51089 |
| Debye temperature 2 | 985.0031 ± 131.7815 |
| Number of oscillators of Einstein mode | 0.75741 ± 0.38378 |
| Einstein temperature | 112.58237 ± 15.19992 |
| $R^2$ | 0.99978 |

**Table S6.** Magnetic vectors.

$T$ = 1.5 K from HB-2A POWDER

| Sites | $M_x$ | $M_y$ | $M_z$ | Modulus ($\mu_B$) |
| --- | --- | --- | --- | --- |
| Co(1) | 0 (Fixed) | -0.91(1) | 2.07(2) | 2.3(1) |
| Co(2) | 0 (Fixed) | 0.91(1) | 1.16(1) | 1.4(1) |

$T$ = 15 K from BL-9 CORELLI

| Sites | Modulus ($\mu_B$) |
| --- | --- |
| Co(1) | 2.388(1) |
| Co(2) | 2.165(1) |

Vectors in Table S9

$T$ = 15 K from HB-2A POWDER

| Sites | Modulus ($\mu_B$) |
| --- | --- |
| Co(1) | 1.65(1) |
| Co(2) | 2.03(1) |

Vectors in Table S10

**Table S7**. Irreducible representations (IR) for propagation vector **q** = (0, 0, 0) and corresponding reduced $\chi^2$ from refinement.

| **MSG** | **IR** | **# Parameters** | **$R_{WP}$** | $\chi^2$ |
|---|---|---|---|---|
| Pnma (62.441) | $\Gamma_1$ | 6 | 15.5 | 2.871 |
| Pn'ma (62.443) | $\Gamma_4$ | 6 | 26.8 | 7.789 |
| Pnm'a (62.444) | $\Gamma_6$ | 6 | 26.1 | 7.356 |
| Pnma' (62.445) | $\Gamma_8$ | 6 | 19.1 | 3.968 |
| Pn'm'a (62.446) | $\Gamma_7$ | 6 | 27.7 | 8.315 |
| Pnm'a' (62.447) | $\Gamma_3$ | 6 | 22.6 | 5.515 |
| Pn'ma' (62.448) | $\Gamma_5$ | 6 | 23.7 | 6.069 |
| Pn'm'a' (62.449) | $\Gamma_2$ | 6 | 25.5 | 7.021 |

**Table S8**. Neutron refinement result of the $\Gamma_1$ irreducible representations in Table S7.

| **Co Type** | **$M_x$** | **$M_y$** | **$M_z$** | **$M_{tot}$** |
|---|---|---|---|---|
| Co(1) | 0 | -0.961 | 2.011 | 2.3(1) |
| Co(2) | 0 | 0.775 | 1.167 | 1.4(1) |

| $\Gamma_1$ | $\psi_1$ | | | $\psi_2$ | | | $\psi_3$ | | |
|---|---|---|---|---|---|---|---|---|---|
| | $m_x$ | $m_y$ | $m_z$ | $m_x$ | $m_y$ | $m_z$ | $m_x$ | $m_y$ | $m_z$ |
| Co(1,2)_1 | 1 | 0 | 0 | 0 | 1 | 0 | 0 | 0 | 1 |
| Co(1,2)_2 | 1 | 0 | 0 | 0 | -1 | 0 | 0 | 0 | -1 |
| Co(1,2)_3 | -1 | 0 | 0 | 0 | 1 | 0 | 0 | 0 | -1 |
| Co(1,2)_4 | -1 | 0 | 0 | 0 | -1 | 0 | 0 | 0 | 1 |
| Co(1,2)_5 | 1 | 0 | 0 | 0 | 1 | 0 | 0 | 0 | 1 |
| Co(1,2)_6 | 1 | 0 | 0 | 0 | -1 | 0 | 0 | 0 | -1 |
| Co(1,2)_7 | -1 | 0 | 0 | 0 | 1 | 0 | 0 | 0 | -1 |
| Co(1,2)_8 | -1 | 0 | 0 | 0 | -1 | 0 | 0 | 0 | 1 |

**Table S9**. Magnetic vectors at $T$ = 14.5 K from BL-9 CORELLI.
Atom : Co(1)

| x | y | z | Translation | m(a) | m(b) | m(c) | Mtot |
|---|---|---|---|---|---|---|---|
| 0.43646 | 0.10306 | 0.39676 | | | | | |
| | | | ( 0, 0, 0) | 0.00000 | -1.58129 | 1.78986 | 2.38832 |
| | | | ( 0, 1, 0) | 0.00000 | 2.01034 | 1.28941 | 2.38832 |
| | | | ( 0, 2, 0) | 0.00000 | 0.96729 | -2.18367 | 2.38832 |
| | | | ( 0, 3, 0) | 0.00000 | -2.30577 | -0.62248 | 2.38832 |
| 0.56354 | 0.60306 | 0.60324 | | | | | |
| | | | ( 0, 0, 0) | 0.00000 | 2.38810 | -0.03250 | 2.38832 |
| | | | ( 0, 1, 0) | 0.00000 | -0.33257 | 2.36505 | 2.38832 |
| | | | ( 0, 2, 0) | 0.00000 | -2.28653 | -0.68983 | 2.38832 |
| | | | ( 0, 3, 0) | 0.00000 | 1.03092 | -2.15436 | 2.38832 |
| 0.93646 | 0.10306 | 0.10324 | | | | | |
| | | | ( 0, 0, 0) | 0.00000 | 1.58129 | 1.78986 | 2.38832 |
| | | | ( 0, 1, 0) | 0.00000 | -2.01034 | 1.28941 | 2.38832 |
| | | | ( 0, 2, 0) | 0.00000 | -0.96729 | -2.18367 | 2.38832 |
| | | | ( 0, 3, 0) | 0.00000 | 2.30577 | -0.62248 | 2.38832 |
| 0.06354 | 0.60306 | 0.89676 | | | | | |
| | | | ( 0, 0, 0) | 0.00000 | -2.38810 | -0.03250 | 2.38832 |
| | | | ( 0, 1, 0) | 0.00000 | 0.33257 | 2.36505 | 2.38832 |
| | | | ( 0, 2, 0) | 0.00000 | 2.28653 | -0.68983 | 2.38832 |
| | | | ( 0, 3, 0) | 0.00000 | -1.03092 | -2.15436 | 2.38832 |
| 0.56354 | 0.89694 | 0.60324 | | | | | |
| | | | ( 0, 0, 0) | 0.00000 | 2.01034 | 1.28941 | 2.38832 |
| | | | ( 0, 1, 0) | 0.00000 | 0.96729 | -2.18367 | 2.38832 |
| | | | ( 0, 2, 0) | 0.00000 | -2.30577 | -0.62248 | 2.38832 |
| | | | ( 0, 3, 0) | 0.00000 | -0.26306 | 2.37379 | 2.38832 |
| 0.43646 | 0.39694 | 0.39676 | | | | | |
| | | | ( 0, 0, 0) | 0.00000 | 2.38810 | -0.03250 | 2.38832 |
| | | | ( 0, 1, 0) | 0.00000 | -0.33257 | 2.36505 | 2.38832 |
| | | | ( 0, 2, 0) | 0.00000 | -2.28653 | -0.68983 | 2.38832 |
| | | | ( 0, 3, 0) | 0.00000 | 1.03092 | -2.15436 | 2.38832 |
| 0.06354 | 0.89694 | 0.89676 | | | | | |
| | | | ( 0, 0, 0) | 0.00000 | -2.01034 | 1.28941 | 2.38832 |
| | | | ( 0, 1, 0) | 0.00000 | -0.96729 | -2.18367 | 2.38832 |
| | | | ( 0, 2, 0) | 0.00000 | 2.30577 | -0.62248 | 2.38832 |
| | | | ( 0, 3, 0) | 0.00000 | 0.26306 | 2.37379 | 2.38832 |
| 0.93646 | 0.39694 | 0.10324 | | | | | |
| | | | ( 0, 0, 0) | 0.00000 | -2.38810 | -0.03250 | 2.38832 |
| | | | ( 0, 1, 0) | 0.00000 | 0.33257 | 2.36505 | 2.38832 |
| | | | ( 0, 2, 0) | 0.00000 | 2.28653 | -0.68983 | 2.38832 |
| | | | ( 0, 3, 0) | 0.00000 | -1.03092 | -2.15436 | 2.38832 |

Atom : Co(2)

| x | y | z | Translation | m(a) | m(b) | m(c) | Mtot |
|---|---|---|---|---|---|---|---|
| 0.07467 | 0.58992 | 0.39579 | | | | | |
| | | | ( 0, 0, 0) | 0.00000 | -1.43317 | 1.62221 | 2.16461 |
| | | | ( 0, 1, 0) | 0.00000 | 1.82204 | 1.16863 | 2.16461 |
| | | | ( 0, 2, 0) | 0.00000 | 0.87668 | -1.97913 | 2.16461 |
| | | | ( 0, 3, 0) | 0.00000 | -2.08980 | -0.56417 | 2.16461 |
| 0.92533 | 0.08992 | 0.60421 | | | | | |
| | | | ( 0, 0, 0) | 0.00000 | -0.35963 | -2.13453 | 2.16461 |
| | | | ( 0, 1, 0) | 0.00000 | 2.16441 | -0.02945 | 2.16461 |
| | | | ( 0, 2, 0) | 0.00000 | -0.30142 | 2.14352 | 2.16461 |
| | | | ( 0, 3, 0) | 0.00000 | -2.07235 | -0.62522 | 2.16461 |
| 0.57467 | 0.58992 | 0.10421 | | | | | |
| | | | ( 0, 0, 0) | 0.00000 | 1.43317 | 1.62221 | 2.16461 |
| | | | ( 0, 1, 0) | 0.00000 | -1.82204 | 1.16863 | 2.16461 |
| | | | ( 0, 2, 0) | 0.00000 | -0.87668 | -1.97913 | 2.16461 |
| | | | ( 0, 3, 0) | 0.00000 | 2.08980 | -0.56417 | 2.16461 |
| 0.42533 | 0.08992 | 0.89579 | | | | | |
| | | | ( 0, 0, 0) | 0.00000 | 0.35963 | -2.13453 | 2.16461 |
| | | | ( 0, 1, 0) | 0.00000 | -2.16441 | -0.02945 | 2.16461 |
| | | | ( 0, 2, 0) | 0.00000 | 0.30142 | 2.14352 | 2.16461 |
| | | | ( 0, 3, 0) | 0.00000 | 2.07235 | -0.62522 | 2.16461 |
| 0.92533 | 0.41008 | 0.60421 | | | | | |
| | | | ( 0, 0, 0) | 0.00000 | 1.82204 | 1.16863 | 2.16461 |
| | | | ( 0, 1, 0) | 0.00000 | 0.87668 | -1.97913 | 2.16461 |
| | | | ( 0, 2, 0) | 0.00000 | -2.08980 | -0.56417 | 2.16461 |
| | | | ( 0, 3, 0) | 0.00000 | -0.23842 | 2.15144 | 2.16461 |
| 0.07467 | 0.91008 | 0.39579 | | | | | |
| | | | ( 0, 0, 0) | 0.00000 | -0.30142 | 2.14352 | 2.16461 |
| | | | ( 0, 1, 0) | 0.00000 | -2.07235 | -0.62522 | 2.16461 |
| | | | ( 0, 2, 0) | 0.00000 | 0.93436 | -1.95257 | 2.16461 |
| | | | ( 0, 3, 0) | 0.00000 | 1.78698 | 1.22158 | 2.16461 |
| 0.42533 | 0.41008 | 0.89579 | | | | | |
| | | | ( 0, 0, 0) | 0.00000 | -1.82204 | 1.16863 | 2.16461 |
| | | | ( 0, 1, 0) | 0.00000 | -0.87668 | -1.97913 | 2.16461 |
| | | | ( 0, 2, 0) | 0.00000 | 2.08980 | -0.56417 | 2.16461 |
| | | | ( 0, 3, 0) | 0.00000 | 0.23842 | 2.15144 | 2.16461 |
| 0.57467 | 0.91008 | 0.10421 | | | | | |
| | | | ( 0, 0, 0) | 0.00000 | 0.30142 | 2.14352 | 2.16461 |
| | | | ( 0, 1, 0) | 0.00000 | 2.07235 | -0.62522 | 2.16461 |
| | | | ( 0, 2, 0) | 0.00000 | -0.93436 | -1.95257 | 2.16461 |
| | | | ( 0, 3, 0) | 0.00000 | -1.78698 | 1.22158 | 2.16461 |

**Table S10**. Magnetic vectors at $T$ = 15 K from HB-2A POWDER

Atom : Co(1)

| x | y | z | Translation | m(a) | m(b) | m(c) | Mtot |
|---|---|---|---|---|---|---|---|
| 0.43646 | 0.10306 | 0.39676 | | | | | |
| | | | ( 0, 0, 0) | 0.00000 | -1.01053 | 1.30964 | 1.65418 |
| | | | ( 0, 1, 0) | 0.00000 | 1.30964 | 1.01053 | 1.65418 |
| | | | ( 0, 2, 0) | 0.00000 | 1.01053 | -1.30964 | 1.65419 |
| | | | ( 0, 3, 0) | 0.00000 | -1.30964 | -1.01053 | 1.65418 |
| 0.56354 | 0.60306 | 0.60324 | | | | | |
| | | | ( 0, 0, 0) | 0.00000 | 1.64061 | 0.21150 | 1.65419 |
| | | | ( 0, 1, 0) | 0.00000 | -0.21150 | 1.64061 | 1.65419 |
| | | | ( 0, 2, 0) | 0.00000 | -1.64061 | -0.21150 | 1.65418 |
| | | | ( 0, 3, 0) | 0.00000 | 0.21150 | -1.64061 | 1.65419 |
| 0.93646 | 0.10306 | 0.10324 | | | | | |
| | | | ( 0, 0, 0) | 0.00000 | 1.01053 | 1.30964 | 1.65418 |
| | | | ( 0, 1, 0) | 0.00000 | -1.30964 | 1.01053 | 1.65418 |
| | | | ( 0, 2, 0) | 0.00000 | -1.01053 | -1.30964 | 1.65419 |
| | | | ( 0, 3, 0) | 0.00000 | 1.30964 | -1.01053 | 1.65418 |
| 0.06354 | 0.60306 | 0.89676 | | | | | |
| | | | ( 0, 0, 0) | 0.00000 | -1.64061 | 0.21150 | 1.65419 |
| | | | ( 0, 1, 0) | 0.00000 | 0.21150 | 1.64061 | 1.65419 |
| | | | ( 0, 2, 0) | 0.00000 | 1.64061 | -0.21150 | 1.65418 |
| | | | ( 0, 3, 0) | 0.00000 | -0.21150 | -1.64061 | 1.65418 |
| 0.56354 | 0.89694 | 0.60324 | | | | | |
| | | | ( 0, 0, 0) | 0.00000 | 1.30964 | 1.01053 | 1.65418 |
| | | | ( 0, 1, 0) | 0.00000 | 1.01053 | -1.30964 | 1.65418 |
| | | | ( 0, 2, 0) | 0.00000 | -1.30964 | -1.01053 | 1.65419 |
| | | | ( 0, 3, 0) | 0.00000 | -1.01053 | 1.30964 | 1.65418 |
| 0.43646 | 0.39694 | 0.39676 | | | | | |
| | | | ( 0, 0, 0) | 0.00000 | 1.64061 | 0.21150 | 1.65419 |
| | | | ( 0, 1, 0) | 0.00000 | -0.21150 | 1.64061 | 1.65419 |
| | | | ( 0, 2, 0) | 0.00000 | -1.64061 | -0.21150 | 1.65418 |
| | | | ( 0, 3, 0) | 0.00000 | 0.21150 | -1.64061 | 1.65419 |
| 0.06354 | 0.89694 | 0.89676 | | | | | |
| | | | ( 0, 0, 0) | 0.00000 | -1.30964 | 1.01053 | 1.65418 |
| | | | ( 0, 1, 0) | 0.00000 | -1.01053 | -1.30964 | 1.65418 |
| | | | ( 0, 2, 0) | 0.00000 | 1.30964 | -1.01053 | 1.65419 |
| | | | ( 0, 3, 0) | 0.00000 | 1.01053 | 1.30964 | 1.65418 |
| 0.93646 | 0.39694 | 0.10324 | | | | | |
| | | | ( 0, 0, 0) | 0.00000 | -1.64061 | 0.21150 | 1.65419 |
| | | | ( 0, 1, 0) | 0.00000 | 0.21150 | 1.64061 | 1.65419 |
| | | | ( 0, 2, 0) | 0.00000 | 1.64061 | -0.21150 | 1.65418 |
| | | | ( 0, 3, 0) | 0.00000 | -0.21150 | -1.64061 | 1.65418 |

Atom : Co(2)

| x | y | z | Translation | m(a) | m(b) | m(c) | Mtot |
|---|---|---|---|---|---|---|---|
| 0.07467 | 0.58992 | 0.39579 | | | | | |
| | | | ( 0, 0, 0) | 0.00000 | -1.23795 | 1.60438 | 2.02646 |
| | | | ( 0, 1, 0) | 0.00000 | 1.60438 | 1.23795 | 2.02646 |
| | | | ( 0, 2, 0) | 0.00000 | 1.23795 | -1.60438 | 2.02646 |
| | | | ( 0, 3, 0) | 0.00000 | -1.60438 | -1.23795 | 2.02646 |
| 0.92533 | 0.08992 | 0.60421 | | | | | |
| | | | ( 0, 0, 0) | 0.00000 | 0.25911 | -2.00983 | 2.02646 |
| | | | ( 0, 1, 0) | 0.00000 | 2.00983 | 0.25911 | 2.02646 |
| | | | ( 0, 2, 0) | 0.00000 | -0.25911 | 2.00983 | 2.02646 |
| | | | ( 0, 3, 0) | 0.00000 | -2.00983 | -0.25911 | 2.02646 |
| 0.57467 | 0.58992 | 0.10421 | | | | | |
| | | | ( 0, 0, 0) | 0.00000 | 1.23795 | 1.60438 | 2.02646 |
| | | | ( 0, 1, 0) | 0.00000 | -1.60438 | 1.23795 | 2.02646 |
| | | | ( 0, 2, 0) | 0.00000 | -1.23795 | -1.60438 | 2.02646 |
| | | | ( 0, 3, 0) | 0.00000 | 1.60438 | -1.23795 | 2.02646 |
| 0.42533 | 0.08992 | 0.89579 | | | | | |
| | | | ( 0, 0, 0) | 0.00000 | -0.25911 | -2.00983 | 2.02646 |
| | | | ( 0, 1, 0) | 0.00000 | -2.00983 | 0.25911 | 2.02646 |
| | | | ( 0, 2, 0) | 0.00000 | 0.25911 | 2.00983 | 2.02646 |
| | | | ( 0, 3, 0) | 0.00000 | 2.00983 | -0.25911 | 2.02646 |
| 0.92533 | 0.41008 | 0.60421 | | | | | |
| | | | ( 0, 0, 0) | 0.00000 | 1.60438 | 1.23795 | 2.02646 |
| | | | ( 0, 1, 0) | 0.00000 | 1.23795 | -1.60438 | 2.02646 |
| | | | ( 0, 2, 0) | 0.00000 | -1.60438 | -1.23795 | 2.02646 |
| | | | ( 0, 3, 0) | 0.00000 | -1.23795 | 1.60438 | 2.02646 |
| 0.07467 | 0.91008 | 0.39579 | | | | | |
| | | | ( 0, 0, 0) | 0.00000 | -0.25911 | 2.00983 | 2.02646 |
| | | | ( 0, 1, 0) | 0.00000 | -2.00983 | -0.25911 | 2.02646 |
| | | | ( 0, 2, 0) | 0.00000 | 0.25911 | -2.00983 | 2.02646 |
| | | | ( 0, 3, 0) | 0.00000 | 2.00983 | 0.25911 | 2.02646 |
| 0.42533 | 0.41008 | 0.89579 | | | | | |
| | | | ( 0, 0, 0) | 0.00000 | -1.60438 | 1.23795 | 2.02646 |
| | | | ( 0, 1, 0) | 0.00000 | -1.23795 | -1.60438 | 2.02646 |
| | | | ( 0, 2, 0) | 0.00000 | 1.60438 | -1.23795 | 2.02646 |
| | | | ( 0, 3, 0) | 0.00000 | 1.23795 | 1.60438 | 2.02646 |
| 0.57467 | 0.91008 | 0.10421 | | | | | |
| | | | ( 0, 0, 0) | 0.00000 | 0.25911 | 2.00983 | 2.02646 |
| | | | ( 0, 1, 0) | 0.00000 | 2.00983 | -0.25911 | 2.02646 |
| | | | ( 0, 2, 0) | 0.00000 | -0.25911 | -2.00983 | 2.02646 |
| | | | ( 0, 3, 0) | 0.00000 | -2.00983 | 0.25911 | 2.02646 |